\documentclass[10pt,a4paper]{article}                                                    

\usepackage{setspace} 

\usepackage[top=2.5cm,bottom=2.5cm,left=1.5cm,right=1.5cm]{geometry}                     

\usepackage{layouts}                                                                     

\usepackage{multicol}                                                                    
\usepackage[small,bf]{caption}                                                           

\usepackage[T1]{fontenc} 
\usepackage{lmodern}     

\usepackage{scalerel}                                                                    

\usepackage{paralist}            
\setdefaultitem{-}{$\triangleright$}{}{}                                                 
\usepackage[ampersand]{easylist} 

\usepackage[usenames,dvipsnames]{xcolor}

\usepackage{amsfonts,amssymb}    
\usepackage[intlimits]{amsmath}  
\usepackage{mathrsfs}            
\usepackage{bm}                  
\usepackage{upgreek}             
\usepackage{dsfont}              
\usepackage{mathtools}           
\usepackage{relsize}             
\usepackage{cancel}              

\DeclarePairedDelimiter\abs{\lvert}{\rvert}%
\DeclarePairedDelimiter\norm{\lVert}{\rVert}%
\makeatletter
\let\oldabs\abs
\def\abs{\@ifstar{\oldabs}{\oldabs*}}
\let\oldnorm\norm
\def\norm{\@ifstar{\oldnorm}{\oldnorm*}}
\makeatother

\usepackage{float}      
\usepackage{subfig}     
\usepackage{booktabs}   

\usepackage{theorem}                                                                     
\theorembodyfont{\normalfont}                                                            
\newtheorem{property}{Property}                                                          

\definecolor{mycc}{rgb}{0, 0, 0.45}
\definecolor{mygreen}{rgb}{0, 0.4, 0}
\usepackage{ifpdf}
\ifpdf
	\usepackage[pdftex=true,
		pdftitle={AdInf},
		pdfauthor={Patrick N. Raanes},
		hyperindex=true,
		colorlinks=true,
		linkcolor=mygreen,
		citecolor=mycc]{hyperref}
	\usepackage{epstopdf}

	\usepackage[stretch=10]{microtype}
\else
	\usepackage[hypertex,
		hyperindex=true,
		colorlinks=false]{hyperref}
	\usepackage{graphicx}
	\usepackage{srcltx}
\fi

\graphicspath{{./}{Pics/}}

\hypersetup{                                                                             
}                                                                                        

\usepackage[colon,authoryear,square]{natbib}                                             
\defcitealias{anderson2007adaptive}{A07}
\defcitealias{anderson2009spatially}{A09}
\defcitealias{miyoshi2011gaussian}{M11}
\defcitealias{li2009simultaneous}{LKM09}
\defcitealias{stroud2007sequential}{SB07}
\defcitealias{sarkka2009adaptiveVB}{Sa09}
\defcitealias{frei2012sequential}{FK12}
\defcitealias{brankart2010efficient}{Br10}
\defcitealias{nakabayashi2017extension}{NU17}
\defcitealias{bocquet2015expanding}{Boc15}
\defcitealias{bocquet2012combining}{Boc12}
\defcitealias{bocquet2011ensemble}{Boc11}

\usepackage{cleveref}
\crefname{equation}{equation}{equations}
\Crefname{equation}{Equation}{Equations}
\crefname{subeqns}{equation}{equations}
\Crefname{subeqns}{Equation}{Equations}
\crefformat{subeqns}{equation~(#2#1#3)}
\Crefformat{subeqns}{Equation~(#2#1#3)}
\labelcrefformat{subeqns}{(#2#1#3)}
\crefname{figure}{Figure}{Figures}
\Crefname{figure}{Figure}{Figures}
\crefname{subfigure}{Figure}{Figures}
\Crefname{subfigure}{Figure}{Figures}
\crefformat{subfigure}{Figure~#2#1#3}
\Crefformat{subfigure}{Figure~#2#1#3}
\crefname{table}{Table}{Tables}
\Crefname{table}{Table}{Tables}
\crefname{theo}{Theorem}{Theorems}
\Crefname{theo}{Theorem}{Theorems}
\crefname{theo}{Theorem}{Theorems}
\Crefname{theo}{Theorem}{Theorems}
\crefname{lemm}{Lemma}{Lemmas}
\Crefname{lemm}{Lemma}{Lemmas}
\crefname{defi}{Definition}{Definitions}
\Crefname{defi}{Definition}{Definitions}
\crefname{prop}{Proposition}{Propositions}
\Crefname{prop}{Proposition}{Propositions}
\crefname{property}{Property}{Properties}
\Crefname{property}{Property}{Properties}
\crefname{coro}{Corollary}{Corollaries}
\Crefname{coro}{Corollary}{Corollaries}
\crefname{algocf}{Algorithm}{Algorithms}
\Crefname{algocf}{Algorithm}{Algorithms}

\newcommand{\NERSC}{Nansen Environmental and Remote Sensing Center}
\newcommand{\NERSCa}{Thorm{\o}hlensgate 47, Bergen, N-5006, Norway}
\newcommand{\CEREA}{CEREA, Joint laboratory of \'Ecole
	des Ponts ParisTech and EDF R\&D,
	Universit\'e Paris-Est, Champs-sur-Marne, France
}

\newcommand{\myemail}{patrick.n.raanes@gmail.com}

\usepackage{authblk}                                                                     
\author[1]{Patrick N. Raanes\thanks{                                                   
   \myemail                                                                              
   }}                                                                                    
\author[2]{Marc Bocquet}                                                                 
\author[1]{Alberto Carrassi}                                                             
\affil[1]{\NERSC, \NERSCa}                                                               
\affil[2]{\CEREA}                                                                        

\title{Adaptive covariance inflation in the ensemble Kalman filter\\
by Gaussian scale mixtures}

\DeclareMathOperator{\diag}{diag}
\DeclareMathOperator{\range}{range}
\DeclareMathOperator{\trace}{tr}
\DeclareMathOperator*{\argmin}{argmin}
\DeclareMathOperator{\image}{image}

\newcommand{\tn}[1]{{\textnormal{{#1}}}}
\newcommand{\Reals}{\mathbb{R}}
\newcommand{\tq}[0]{\; : \;}

\DeclareMathAlphabet{\mathpzc}{OT1}{pzc}{m}{it}
\newcommand{\tdist}{\mathop{}\! {{{\text{\Large$\mathpzc{t}$}}}}}

\newcommand{\Expect}[0]{\mathop{}\! \mathbb{E}}
\newcommand{\NormDist}{\mathop{}\! \mathcal{N}}
\newcommand{\InvWish}{\mathop{}\! \mathcal{W}^{-1}}
\newcommand{\Wishart}{\mathop{}\! \mathcal{W}^{+1}}
\newcommand{\WishartN}{\mathop{}\! \mathcal{W}}
\newcommand{\InvCS}{\mathop{}\! {{{\text{\Large$\mathpzc{\chi}$}}}^{-2}}}
\newcommand{\CS}{\mathop{}\! {{{\text{\Large$\mathpzc{\chi}$}}}^{+2}}}
\newcommand{\CSN}{\mathop{}\! {{{\text{\Large$\mathpzc{\chi}$}}}}}
\newcommand{\CWish}{c_{\mathcal{W}}}
\newcommand{\CNorm}{c_{\mathcal{N}}}
\newcommand{\Ctdist}{c_{{{\text{\normalsize$\mathpzc{t}$}}}}}
\newcommand{\CCS}{c_{ {{\text{\normalsize$\mathpzc{\chi}$}}}}}
\newcommand{\CdfNorm}{F_{\NormDist}}
\newcommand{\CdfCS}{ F_{{{ \text{\normalsize$\mathpzc{\chi}$} }}} }
\newcommand{\CovSpace}{\mathcal{B}}
\newcommand{\RestSpace}{\mathcal{C}}
\newcommand{\dynm}[0]{\mathcal{M}}
\newcommand{\dynLin}{\dynm_{\tn{Lin}}}
\newcommand{\dynNonL}{\dynm_{\tn{NonLin}}}
\newcommand{\obsm}[0]{\bH}
\newcommand{\obsmf}[1]{\bH{#1}}

\newcommand{\trsign}{{\mathsf{T}}}
\newcommand{\Sqrtsign}{{1/2}}
\newcommand{\tr}{\ensuremath{^{\trsign}}}
\newcommand{\mtrsqrt}{\ensuremath{^{{-\trsign/2}}}}

\newcommand{\sq}{\ensuremath{^\Sqrtsign}}
\newcommand{\msq}{\ensuremath{^{-\Sqrtsign}}}
\newcommand{\eN}[0]{\varepsilon_N}
\newcommand{\ones}[0]{\mathds{1}}

\newcommand*{\pdf}{\mathop{}\! p}
\newcommand*{\qdf}{\mathop{}\! q}
\newcommand*{\diff}{\mathop{}\!\mathrm{d}}

\newcommand{\dd}[2]{\frac{\diff #1}{\diff #2}}
\newcommand{\pd}[2]{\frac{\mathrm{\partial}{#1}}{\mathrm{\partial}{#2}}}

\newcommand{\mahalf}[0]{{\textstyle{-\frac{1}{2}}}}

\newcommand{\iid}[0]{iid}
\newcommand{\PDF}[0]{pdf}
\newcommand{\vicev}[0]{vice versa}

\newcommand{\apost}[0]{a posteriori}

\newcommand{\bvec}[1]{{\bm{#1}}}
\newcommand{\x}[0]{\bvec{x}}
\newcommand{\w}[0]{\bvec{w}}
\newcommand{\e}[0]{\bvec{e}}
\newcommand{\y}[0]{\bvec{y}}
\newcommand{\z}[0]{\bvec{z}}
\newcommand{\bb}[0]{\bvec{b}}
\newcommand{\bu}[0]{\bvec{u}}
\newcommand{\bv}[0]{\bvec{v}}
\newcommand{\up}[0]{\bvec{\upsilon}}
\newcommand{\Q}[0]{\mat{Q}}
\newcommand{\bPi}[0]{\mat{\Pi}}
\newcommand{\PiAN}[0]{\bPi_\ones^\perp}
\newcommand{\bx}[0]{\bvec{\bar{x}}}
\newcommand{\bw}[0]{\bvec{\bar{w}}}
\newcommand{\bxi}[0]{\bvec{\xi}}
\newcommand{\bdelta}[0]{\bvec{\delta}}
\newcommand{\bbdelta}[0]{\bvec{\bar{\delta}}}

\newcommand{\mat}[1]{{\mathbf{{#1}}}}
\newcommand{\barK}[0]{\mat{\bar{K}}}

\newcommand{\barPa}[0]{\mat{\bar{P}}^\tn{a}}

\newcommand{\barB}[0]{\mat{\bar{B}}}

\newcommand{\barC}[0]{\mat{\bar{C}}}

\newcommand{\X}[0]{\mat{X}}
\newcommand{\Y}[0]{\mat{Y}}
\newcommand{\E}[0]{\mat{E}}
\newcommand{\U}[0]{\mat{U}}
\newcommand{\V}[0]{\mat{V}}

\newcommand{\T}[0]{\mat{T}}
\newcommand{\C}[0]{\mat{C}}
\newcommand{\bS}[0]{\mat{S}}
\newcommand{\I}[0]{\mat{I}}

\newcommand{\bP}[0]{\mat{P}}
\newcommand{\bH}[0]{\mat{H}}
\newcommand{\R}[0]{\mat{R}}
\newcommand{\B}[0]{\mat{B}}
\newcommand{\M}[0]{\mat{M}}

\newcommand{\compactN}[0]{{N{-}1}}
\newcommand{\cN}[0]{(\compactN)}
\newcommand{\fracN}[0]{{{\tfrac{1}{N-1}}}} 
\newcommand{\AN}[0]{{\big( \I_N - \ones\ones\tr /N \big)}}
\newcommand{\RMSE}[0]{{\tn{RMSE}}}

\newcommand{\scl}[0]{{\tilde{\alpha}}}

\newcommand{\approptoinn}[2]{\mathrel{\vcenter{
 \offinterlineskip\halign{\hfil$##$\cr
	#1\propto\cr\noalign{\kern1.3pt}#1\sim\cr\noalign{\kern-2pt}}}}}
\newcommand{\appropto}{\mathpalette\approptoinn\relax}

\begin{document}
\newcommand{\makeAbstractCmd}[1]{
	\newcommand*{\myabstract}{
		\begin{abstract}
			#1
		\end{abstract}
	}
}
\makeAbstractCmd{
		This paper studies multiplicative inflation:
		the complementary scaling of the state covariance in the ensemble Kalman filter (EnKF).
		Firstly, error sources in the EnKF are catalogued and discussed in relation to inflation;
		nonlinearity is given particular attention as a source of sampling error.
		In response,
		the ``finite-size'' refinement known as the EnKF-$N$ is re-derived
		via a Gaussian scale mixture,
		again demonstrating how it yields adaptive inflation.
		Existing methods for adaptive inflation estimation are reviewed,
		and several insights are gained from a comparative analysis.
		One such adaptive inflation method is selected to complement the EnKF-$N$ to make a hybrid that is
		suitable for contexts where model error is present and imperfectly parameterized.
		Benchmarks are obtained from experiments with the two-scale Lorenz model
		and its slow-scale truncation.
		The proposed hybrid EnKF-$N$ method of adaptive inflation is found to yield systematic accuracy improvements
		in comparison with the existing methods,
		albeit to a moderate degree.
}

\newcommand*{\draftV}{
\begin{center}
	\vspace{-2em}
	Draft version 16
\end{center}
}


\twocolumn[                                                                              
  \begin{@twocolumnfalse}                                                                
    \maketitle                                                                           
    \vspace{-1.4em}                                                                      
    \myabstract~\\                                                                       
  \end{@twocolumnfalse}                                                                  
]                                                                                        



\section{Introduction}
\label{sec:Introduction}

Consider the problem of
estimating the state $\x_k \in \Reals^M$
given the observation $\y_k \in \Reals^P$,
as generated by:
\begin{subequations}
	\begin{align}
		\label[equation]{eqn:state_sde_discrete}
		\x_{k} &= \dynm(\x_{k-1}) + \bxi_k \, ,  && \bxi_k \sim \NormDist(\bvec{0},\Q_k) \, , \\
		\label[equation]{eqn:obs_eqn}
		\y_k     &= \obsmf{\x_k} + \up_k \, ,  && \up_k \sim \NormDist(\bvec{0},\R_k) \, ,
	\end{align}
	\label[subeqns]{eqn:HMM}%
\end{subequations}
for sequentially increasing time index $k$,
where the Gaussian noise processes, $\bxi_k$ and $\up_k$,
are independent in time and from each other.
More specifically, 
the Bayesian filtering problem consists of computing and representing $\pdf(\x_k|\y_{1:k})$,
namely the probability density function (\PDF{})
of the current state, $\x_k$, given the current and past observations,
$\y_{1:k} = \{\y_l\}_{l=1}^k$.
%
In data assimilation (DA) for the geosciences,
the state size, $M$, and possibly the observation size, $P$, may be large,
and the dynamical operator, $\dynm$, may be nonlinear
(observation operators that are nonlinear are implicitly included
by state augmentation \citep{evensen2003ensemble}).
These difficulties necessitate approximate solution methods such as the ensemble Kalman filter (EnKF),
which is simple and efficient \citep{evensen2009ensemble}.

The EnKF computes an ensemble of $N$ realizations, or ``members'',
to represent $\pdf(\x_k|\y_{1:k})$ as a (supposed) sample thereof.
It consists of a forecast-analysis ``cycle''
for each sequential time window of the DA problem.
The forecast step simulates the dynamical forecast
\labelcref{eqn:state_sde_discrete} for each individual member.
This paper is focused on the analysis step.
Since the analysis only concerns a fixed time, $k$,
this subscript is henceforth dropped,
as is the explicit conditioning on 
$\y_{1:k-1}$.
Thus, the prior at time $k$ is written $\pdf(\x)$,
and the analysis (posterior) at time $k$ becomes $\pdf(\x|\y) \propto \pdf(\y|\x) \pdf(\x)$,
per Bayes' rule.

Denote $\{\x_n\}_{n=1}^N$ the forecasted ensemble
representing $\pdf(\x)$,
and define the prior sample mean and covariance:
\begin{subequations}
	\begin{align}
			\label[equation]{eqn:bx}
			\bx   &= \frac{1}{N} \sum_{n=1}^N \x_n\, ,&
			\\
			\label[equation]{eqn:barB}
			\barB &= \frac{1}{N-1} \sum_{n=1}^N \left( \x_n - \bx \right)
			\left( \x_n - \bx \right)\tr \, .
	\end{align}
	\label[subeqns]{eqn:bx_bP_def_0}%
\end{subequations}
The EnKF analysis update can be derived by assuming that
$\bx$ and $\barB$ exactly equal the \emph{true} moments of $\pdf(\x)$,
labelled $\bb$ and $\B$,
and carefully dealing with rank issues \citep[\S 6.2 of][]{raanes2016thesis}.
The posterior then arises as in the Kalman filter,
described by the analysis moments $\bx^\tn{a}$ and $\barPa$,
or a (deterministic, ``square-root'') ensemble transformation to match these.

Multiplicative inflation is an auxiliary technique to adjust (typically increase)
the ensemble spread and thereby covariance, initially studied by
\citet{pham1998singular,anderson1999monte,hamill2001distance}.
Here, the specific variant studied is that of
multiplying the prior state covariance matrix, $\barB$,
by the inflation factor, $\alpha>0$, ahead of the analysis:
\begin{align}
	\barB \mapsto \alpha \barB
	\, .
\end{align}

The need for inflation may arise from intrinsic deficiencies of the EnKF:
errors due to non-Gaussianity or the finite size of the ensemble.
The technique of localization should be applied as the primary remedy,
but inflation is still generally necessary and beneficial \citep[][figure 6.6]{asch2016data}.
Inflation may also be necessary as a heuristic but pragmatic treatment for
extrinsic deficiencies, i.e. model and observational errors,
meaning any misspecification of \cref{eqn:state_sde_discrete,eqn:obs_eqn}.
Again, however, it is advisable to exploit any prior knowledge of errors
(bias, covariance, subspace, etc.)
with more advanced treatments before employing multiplicative inflation.
Examples include additive noise \citep{whitaker2012evaluating},
relaxation \citep{kotsuki2017adaptive},
and square-root transformations \citet{raanes2014ext,sommer2018additive}.

It is difficult to formulate directives for the tuning configurations of the EnKF with any generality.
Concerning $\alpha$,
it may be that the accuracy of the EnKF is improved either by
well-tuned inflation ($\alpha>1$) or deflation ($\alpha<1$).
For example,
as detailed in \cref{sec:catalogue}, sampling error promotes the use of inflation. 
By contrast, the consequences of non-Gaussianity are less transparent.
Nevertheless, it generally seems reasonable to inflate
because non-Gaussianity yields an error (intrinsic to the EnKF)
\emph{adding} to other errors.
Similarly, inflating is typically required in conditions of
extrinsic error such as model error
\citep{li2009simultaneous}.

Further specificity and quantitative guidelines are difficult to deduce.
Therefore, the inflation parameter typically requires application-specific,
off-line tuning for a fixed value,
sometimes at significant expense.
As an alternative strategy,
adaptive inflation aims to estimate the inflation factor on-line.
This also naturally promotes the use of time-varying values.

The EnKF-$N$
\citep[][hereafter \citetalias{bocquet2015expanding}]{bocquet2015expanding}
is a refinement of the analysis step of the EnKF
that explicitly accounts for sampling error in $\bx$ and $\barB$,
meaning their discrepancy from the true moments $\bb$ and $\B$,
which are seen as uncertain, hierarchical ``hyperparameters''.
The derivation proceeds from the rejection of the assumption that $\bx$ and $\barB$ are exact
\citep[][hereafter \citetalias{bocquet2011ensemble}]{bocquet2011ensemble}.
Moreover, when using a non-informative hyperprior for $\bb$ and $\B$,
the EnKF-$N$ has been shown to yield a ``dual'' form which
can be straightforwardly identified as a scheme for adaptive inflation
\citep[][]{bocquet2012combining}.
Its implementation only requires minor add-ons to the (square-root) EnKF,
with negligible computational cost.
In the idealistic context
where model error is absent or accurately parameterized by the noise process,
as detailed by \cref{sec:ideal_context},
the EnKF-$N$ nullifies the need for inflation tuning,
making it opportune for synthetic experiments.
However, (i) wider adoption of the EnKF-$N$ has been limited by some technically challenging aspects of its derivation.
Moreover, (ii) the idealism of the above context means that the EnKF-$N$ would still be reliant on
ad-hoc inflation tuning in real-world, operational use.


This paper addresses both of the above issues of the EnKF-$N$.
Firstly, by re-deriving it with a focus on inflation,
\cref{sec:Re-deriv} further elucidates its workings.
Then, \cref{sec:lit_review_0} reviews and analyses the literature on adaptive inflation estimation.
In contrast to the EnKF-$N$, these adaptive inflation methods have hyperpriors that are
time-dependent (as opposed to being ``reset'' at each analysis time) making them suitable
for realistic contexts where model error is present and imperfectly parameterized.
Then, \cref{sec:hybrid} uses one such method to complement the EnKF-$N$
and create a new, hybrid method.
Lastly, \cref{sec:Benchmark_experiments} presents benchmark experiments
of the various adaptive inflation methods. 
Expressions and properties of the standard parametric \PDF{}s
in use in this paper,
$\NormDist, \tdist, \CS, \InvCS, \Wishart, \InvWish$,
can be found in \cref{sec:standard_pdfs}.

\section{Idealistic contexts and sampling error in the EnKF}
\label{sec:ideal_context}

Model-error adaptive inflation is considered from \cref{sec:lit_review_0} and onward.
By contrast, this section is focused on the effects of sampling error,
as well as its causes, especially nonlinearity.
\Cref{sec:Re-deriv} will show how sampling error is partially remedied by the EnKF-$N$.

\subsection{Two univariate experiments}
\label{sec:Two_scalar_experiments}
Consider the univariate (scalar) filtering problem where
the likelihood $\pdf(y|x) = \NormDist(0|x,2)$
and dynamical model $\dynLin(x) = \sqrt{2} x$
repeat identically for each time index,
and the initial prior is $\pdf(x) = \NormDist(x|0,2)$.
This is a computational (rather than estimation) problem for the posterior;
it is highly artificial, with its numeric values set so as to yield a simple solution.
Indeed, as is perfectly computed by the Kalman filter,
the initial posterior is then $\pdf(x|y) = \NormDist(x|0,1)$,
yielding a forecast prior that is identical to the initial prior.
The cycle thus repeats identically through time.

Now consider the same problem except with nonlinear dynamics,
$\dynNonL(x)$, detailed in \cref{sec:MNonlin}.
This model has been designed to preserve Gaussianity
despite being nonlinear: if $\pdf(x|y) = \NormDist(x|0,1)$
then $\dynNonL(x)$ has the distribution $\NormDist(0,2)$.
Hence the nonlinear DA problem has
exactly the same solution as the linear one.

\begin{figure}[htbp]
	\centering
	\includegraphics[width=0.48\textwidth]{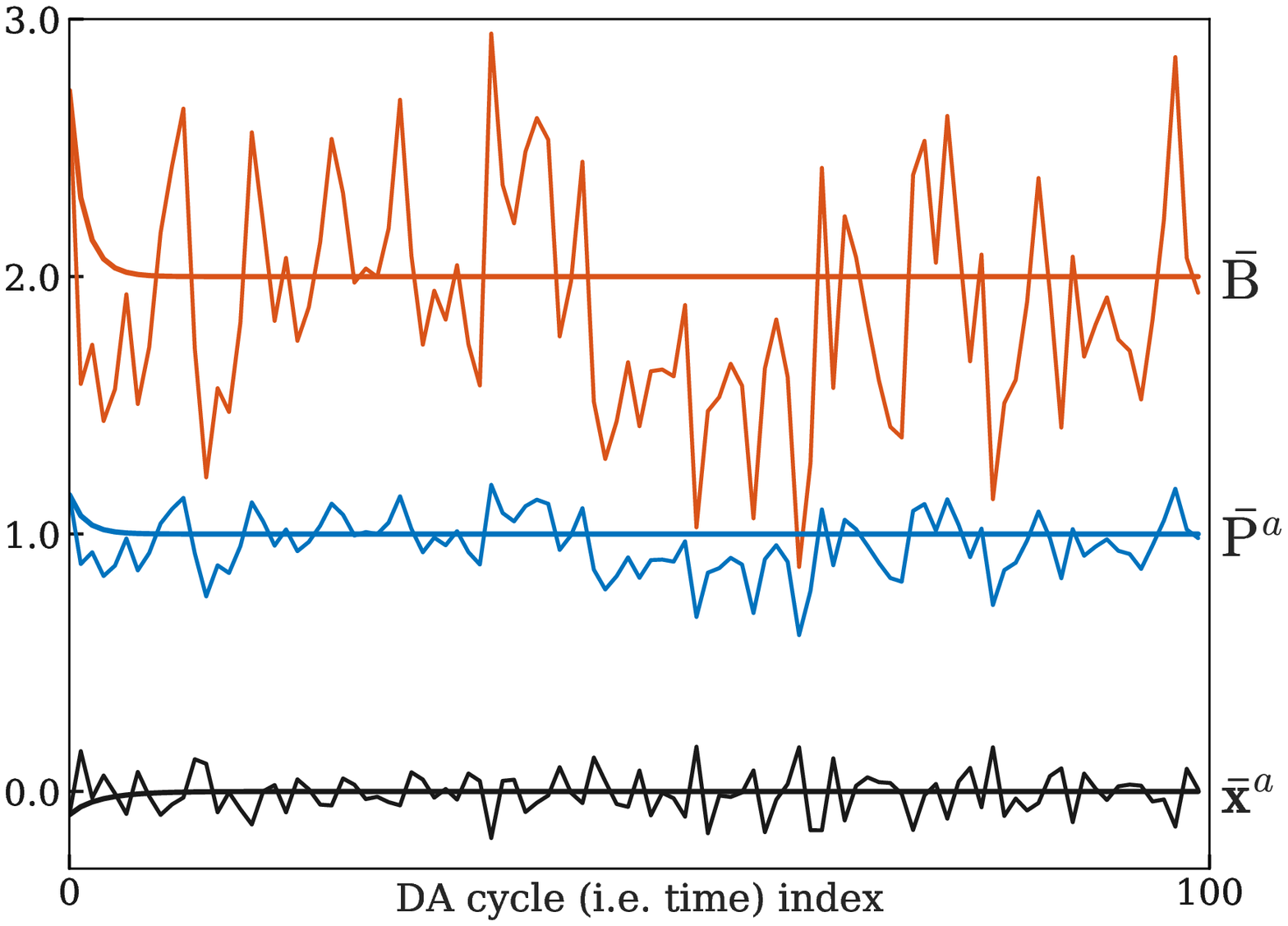}
	\caption[DA experiment illustrating that sampling error is caused by nonlinearity.]
	{
		Time series of statistics from the EnKF applied to the univariate DA problem
		with $\dynLin$ (smooth lines)
		and $\dynNonL$ (jittery lines).
	}
	\label{fig:NonLin_yet_Gaussian}
\end{figure}

However,
as illustrated in \cref{fig:NonLin_yet_Gaussian},
applying a deterministic square-root EnKF (without any inflation or other fixes) to the two problems
yields significantly contrasting results.
The initial ensemble is identical for both cases,
consisting of $N=40$ members drawn randomly from $\pdf(x)$.
But, in the linear case, the resulting sampling errors are quickly attenuated,
and the ensemble statistics converge to the exact ones.

By contrast, in the nonlinear case, the jitteriness (sampling error) is chronic.
This demonstrates that sampling error may arise purely due to nonlinearity,
i.e. without actual stochasticity.
Furthermore, note that the true distributions are perfectly Gaussian,
and therefore the EnKF would compute the exact solution if $N$ were infinite.
Thus, even though nonlinearity typically yields non-Gaussianity, this is not always the case.
Hence, the issue of sampling error, even if caused by nonlinear models,
can be analysed and addressed separately from the issue of non-Gaussianity.

An instructive scenario (not shown) of the nonlinear experiment is that in which
the initial ensemble has a mean of $0$ and a variance of $2$, exactly.
Despite the ``perfect'' initialization,
sampling errors will still be generated,
as predicted by \cref{sec:Why_NSE}.
However, this error is not immediately as big as if
the ensemble were actually randomly sampled from $\NormDist(0,2)$,
in which case $\barB \sim \CS(2,N{-}1)$,
and its expected squared error is $\Expect[\barB - 2]^2 = 8/\cN$,
per \cref{tab:pdfs}.
Indeed, repeated experiments indicate that it takes about
5 consecutive applications of $\dynNonL$
for the ensemble to saturate at a noise level of $8/\cN$.
This gradual build-up also reflects the rule-of-thumb
that stronger nonlinearity breeds larger sampling error.

\subsection{Cataloguing the circumstances for inflation}
\label{sec:catalogue}
This subsection is summarized in \cref{tab:sampling_err_v2},
whose rows correspond to paragraphs, as numbered (\S).

\begin{table}[h]
	\caption{Summary of \cref{sec:catalogue} regarding
		filtering contexts and the consequent need for inflation.
		Background assumptions are idealistic: $\dynm, \obsm, \Q, \R$ perfectly known,
		and $\pdf(\x)$ and $\pdf(\y|\x)$ always Gaussian.
		The star (*) means ``in either case''.
		\vspace{-3ex}}
		\label{tab:sampling_err_v2}
	\begin{center}
	\begin{tabular}[h]{l|lll|l}
		\toprule
		    & Ensemble         & Treatment of      & Model             & Should          \\
	  \S  & size ($N$)       & noises ($\Q, \R$) & ($\dynm$)         & inflate?        \\
		\midrule
		1   & $ \infty       $ & *                        & *                 & No              \\
		2   & $ (M,\infty)   $ & Stochastic               & *                 & Yes             \\
		3   & $ (M,\infty)   $ & *                        & Nonlin.           & Yes             \\
		4   & $ (M,\infty)   $ & Deterministic            & Linear            & No              \\
		5   & $ [2, M]       $ & *                        & *                 & Yes             \\
		\bottomrule
	\end{tabular}
	\end{center}
\end{table}
\newcommand{\nonG}[0]{\cancel{\NormDist}}

\S 1.\hspace{0.5em}%
An important property of the EnKF is that
it is a consistent estimator in the linear-Gaussian case
\citep{tran2009large,mandel2011convergence}:
at each time $k$,
the EnKF statistics $\bx$ and $\barB$
converge (in probability, as $N \rightarrow \infty$)
to the true moments, $\bb$ and $\B$.
Clearly, in this context, inflating or deflating will degrade the ensemble estimates.

\S 2.\hspace{0.5em}%
Stochastic forms of the EnKF employ pseudo-random ``observation perturbations''
for the analysis update step.
Similarly, the forecast step may simulate additive or more advanced stochastic parameterizations
of the forecast noise.
With $N<\infty$, this introduces sampling error.

One cause of the typical need for $\alpha>1$ is the negative bias
of the posterior ensemble covariance matrix \citep{van1999comment,snyder2012intro}:
\begin{align}
	\label{eqn:EnKF_bias_3}
	\Expect[\trace(\barPa)] < \trace(\bP^\tn{a})
	\, ,
\end{align}
where the expectation is taken over the prior ensemble
(or equivalently the covariance, $\barB$), and
\begin{align}
	\barPa &= (\barB^{-1} + \bH\tr \R^{-1} \bH)^{-1} \ ,
	\\
	\bP^\tn{a} &= (\B^{-1} + \bH\tr \R^{-1} \bH)^{-1} \, .
\end{align}
In other words, even though $\Expect[\barB] = \B$,
the nonlinearity (concavity) of $\barPa$,
as a function of $\barB$,
causes a bias.
A related but distinct bias applies for the Kalman gain matrix,
$\barK = \barB\bH\tr (\bH\barB\bH\tr + \R)^{-1}$.
Note, though, that the sampling error originates in the prior;
therefore, the prior covariance is the root cause,
and targeting (inflating) $\barB$,
rather than $\barPa$ and $\barK$,
is more principled.

There is a misconception
that this bias leads to ensemble ``collapse'',
meaning that $\barPa \rightarrow \mat{0}$ and $\barB \rightarrow \mat{0}$ as $k \rightarrow \infty$.
But no matter how acute the single-cycle bias is,
its accumulation will saturate,
because it is counteracted by reductions in $\barK$.

The term ``inbreeding'' is sometimes used
to refer to the bias \labelcref{eqn:EnKF_bias_3}.
However, inbreeding also encompasses two other issues,
namely the introduction of non-Gaussianity and of dependency between ensemble members.
These are caused by the cross-member interaction
that takes place through the EnKF update \citep{houtekamer1998data}.
It is not quite clear how these effects will
impact the need for inflation in later cycles.

Analytical, quantitative results on the bias \labelcref{eqn:EnKF_bias_3}
have been obtained for the general, multivariate case by
\citet{furrer2007estimation,sacher2008sampling}.
However, the degree of the approximation is not entirely clear,
the assumption of the ensemble being truly stochastic is unreliable,
and the related correctional methods were only moderately successful.
An alternative approach is that of {\S 15.3} of \citet[][]{evensen2009data},
where the bias is empirically estimated by using a companion ensemble of white noise.

However, as discussed below \cref{eqn:post_EB},
a significant drawback of the inflation methods targeting this bias
is that they do not establish a feedback mechanism through the cycling of DA.
Moreover, as shown by the theory of the EnKF-$N$ in \cref{sec:Re-deriv},
even in a single cycle, the observations, $\y$, contain information that can improve
estimates of prior hyperparameters ``before'' utilising $\y$ to update the state vector, $\x$,
thereby reducing sampling error and biases.

\S 3.\hspace{0.5em}%
Deterministic, square-root update forms of the EnKF
(which may also be formulated for the forecast noise \citep{raanes2014ext})
do not introduce sampling error in the mean and covariance.
Yet, with $N<\infty$, sampling errors will arise due to model nonlinearities.
This was illustrated in the experiments of \cref{sec:Two_scalar_experiments},
and predicted by \cref{sec:Why_NSE}.
As in {\S 2}, sampling error will instigate the need for inflation.
%
Indeed, the bias \labelcref{eqn:EnKF_bias_3}
is slightly visible in the nonlinear experiment
of \cref{fig:NonLin_yet_Gaussian},
where the covariances, $\barB$ and $\barPa$, are on average lower
(long-run averages: $1.95$ and $0.98$) than the true values.

Filter ``divergence'' is the situation where
the actual error is far larger than expected from $\barB$.
It cannot occur in the linear context,
except by extrinsic errors \citep[][]{fitzgerald1971divergence}.
It may, however, arise in nonlinear, chaotic contexts because,
heuristically,
(i) smaller covariances are prone to deficient (relative) growth by the forecast,
creating an instability that
(ii) might not be adequately controlled by the analyses.
Further, the deficiency in growth typically depends on the starting deficiency of the covariance,
a form of positive feedback that makes the cycle even more ``vicious''.
The alarming prospect of divergence,
especially in light of the bias \labelcref{eqn:EnKF_bias_3},
favours ``erring on the side of caution'', i.e. using $\alpha>1$.

\S 4.\hspace{0.5em}%
With a deterministic, square-root EnKF in the linear context,
sampling error can only come from the initial ensemble and,
as was observed in the experiments of \cref{sec:Two_scalar_experiments},
it will be attenuated through the filtering cycles.
Thus, except perhaps from an initial transitory period,
it is not advisable to use inflation.
This is not always true in experiments, however,
because numerical instabilities (or countermeasures such as regularization)
may allow for improved accuracy with some inflation.

The attenuation of sampling errors can be explained as follows.
Apart from the erroneous initial covariance,
the square-root EnKF is here analytically equivalent
to the Kalman filter \citep{bocquet2016aus4d}.
Thus, the covariance obeys the Riccati recurrence,
which forgets its initial (erroneous) condition,
also in the case of $\Q=\mat{0}$ \citep{bocquet2017degenerate}.
Hence, convergence (in time $k$) holds for any $N \geq M$,
with a rate independent of $N$.

Interestingly, a similar analysis reveals that
the choice of normalization factor for the covariance estimator
(usually $\frac{1}{N-1}$, or $\frac{1}{N}$)
does not impact the asymptotic EnKF-estimated moments (in the linear context):
they always converge to the true moments as $k\rightarrow \infty$.
This means that the success of the EnKF
does not so much rely on some statistical, single-cycle optimality
or unbiasedness (in $\barB$, $\barPa$, or $\barK$),
but rather on the above insensitivity to the choice of normalization factor.

\S 5.\hspace{0.5em}%
Decreasing the ensemble size, $N$,
increases the sampling error,
the bias \labelcref{eqn:EnKF_bias_3},
and the need for inflation.
Worse, if $N \leq M$, then the ensemble is said to be rank-deficient;
this is a separate issue from sampling error,
with the grave consequence that
the truth, $\x$, will not lie entirely within the ensemble subspace (cf. \cref{sec:CVar}).
By operating marginally,
``localization'' \citep{anderson2003local,sakov2011relation},
can mitigate the rank deficiency.
Localization also diminishes off-diagonal sampling errors (``spurious correlations''),
thus decreasing the need for inflation.
On the other hand, by eliminating prior correlations,
localization affects an overly uncertain prior,
yielding too strong a reduction of the ensemble spread\footnote{%
	Formally, quantify the reduction via $|\I_P - \bH \barK| = |\R| / | \bH \barB \bH\tr + \R|$,
	the determinant of the reduction in the variance.
	Localization decreases the magnitude of the off-diagonals of $\bH \barB \bH\tr + \R$,
	provided the eigen-structures of the two terms are not too dissimilar.
	Thus, localization increases the denominator,
	hence reducing $\I_P - \bH \barK$ and the posterior variance.
}.

Another consequence of rank deficiency
is the possibility of the Bayesian uncertainty (i.e. potential error)
outside of the ensemble subspace ``mixing in'', and adding to, the ensemble subspace uncertainty.
If $\Q = \mat{0}$ and the context is linear,
this interaction is small and transitory.
It then does not seem beneficial to (inflate in order to)
have the ensemble spread match the total (as opposed to the subspace) uncertainty.
By contrast, if $\Q>\mat{0}$ \citep{grudzien2018chaotic},
or in the nonlinear context \citep{palatella2015interaction},
the interaction will occur, favouring the use of $\alpha>1$.
In their section 4, \citetalias[][]{bocquet2015expanding} showed that
(scalar/homogeneous) inflation is well-suited to combat this type of error;
this applies for both multiplicative and additive treatments.

Assuming $\Q = \mat{0}$, the long-run ($k \rightarrow \infty$) rank of the true state covariance, $\B$,
is the number of non-decaying modes (non-negative Lyapunov exponents) of the dynamics,
i.e. the rank of the ``unstable subspace'', $0 \leq n_0 < M$.
This correspondence also holds approximately in the nonlinear context,
and means that the rank deficiency of the ensemble may be much less severe than $M-N+1$
\citep{bocquet2016aus4d}.
If this is the case, a duplicate of \cref{tab:sampling_err_v2} applies,
with $M$ replaced by $n_0$.

Filter divergence will (almost surely) occur if $N \leq n_0$,
if localization is not used.
In contrast with \S 3,
inflation is then futile,
because the divergence is caused by rank deficiency,
regardless of the degree of nonlinearity of the growth.
It could be speculated that nonlinearity will
sequentially ``rotate'' the ensemble around in the unstable subspace,
and hence effectively encompass it.
However, twin experiments with the 40-dimensional Lorenz model,
such as the data point $N=n_0=14$ of Figure 6.6 of \citep[][]{asch2016data},
do not give credence to this hypothesis.

\section{Re-deriving the dual EnKF-$N$ via a Gaussian scale mixture}
\label{sec:Re-deriv}
This \namecref{sec:Sketch} gives a new derivation of the dual EnKF-$N$.
Subsection \labelcref{sec:Sketch} outlines the main ideas.
The details are filled in by the subsequent subsections.

\subsection{Overview of the derivation}
\label{sec:Sketch}
Suppose
the Bayesian forecast prior for the ``truth'' is Gaussian,
with mean $\bb$ and covariance $\B$;
formally,
$\pdf(\x | \y_{1:k-1}) = \NormDist(\x | \bb, \B)$,
where the conditioning on past observations has been made explicit again.
Furthermore, assume that the sample $\{\x_n\}_{n=1}^N$ is an ``ensemble'',
meaning that its members are independent and
statistically indistinguishable from the truth \citep{wilks2011statistical},
having been drawn from the very same distribution.
In short,
\begin{align}
	\label{eqn:const_4}
	\x \text{ and } \x_n \sim \NormDist(\bb, \B) \text{\textit{ \iid{}}.}
\end{align}
The assumption \labelcref{eqn:const_4} is convenient,
but may be too idealistic in case of severe
inbreeding, non-Gaussianity, and model error.
Conversely, it may be too agnostic in case the ensemble is not fully random,
as discussed in \cref{sec:Two_scalar_experiments}.
For convenience, assemble the ensemble into the matrix
$\E = \begin{bmatrix} \x_1, & \ldots & \x_n, & \ldots & \x_N \end{bmatrix}$.

Even in the linear-Gaussian context,
computational constraints induce the use of an ensemble to carry the information on the state,
and thus the approximation
\begin{align}
	\pdf(\x | \y_{1:k-1}) \approx \pdf(\x|\E) \, ,
\end{align}
meaning the reduction of the information of $\y_{1:k-1}$ to that represented by the forecast ensemble, $\E$.
Thus, while in principle (with infinite computational resources)
the ``true moments'', $\bb$ and $\B$, are known,
this is not so when employing the EnKF.
Here, all that is known about $\bb$ and $\B$ comes from $\E$.

The appropriate response is to consider all of the possibilities;
indeed, since by the above assumptions
$ \pdf (\x, \bb, \B | \E) = \NormDist (\x | \bb, \B ) \pdf (\bb, \B | \E)$,
marginalization yields:
\begin{align}
	\pdf(\x|\E)
	&=
	\int_\CovSpace \int_{\Reals^{M}}
	\NormDist (\x | \bb, \B)
	\pdf (\bb, \B | \E) \diff \bb \diff \B
	\, ,
	\label{eqn:hier_5}
\end{align}
where $\CovSpace$ is the set of $M {\times} M$ (symmetric) positive-definite matrices\footnote{%
	$\CovSpace$ is the Euclidean space $\Reals^{M(M+1)/2}$
	corresponding to the $M(M+1)/2$ upper-triangular elements in $\B$,
	restricted to positive-definite matrices (the conic subset wherein $\B > \mat{0}$).
}.
\Cref{eqn:hier_5} says that the ``effective prior'',
$\pdf(\x|\E)$,
is a (continuous) mixture:
the average of the ``candidate priors'',
$\NormDist (\x | \bb, \B)$,
as weighted by the ``mixing distribution'',
$\pdf (\bb, \B | \E)$.
Since the distribution of the state, $\x$, depends on
the abstract parameters $\bb$ and $\B$ that are themselves unknown,
these are called hyperparameters and this layered structure is called hierarchical.

The standard EnKF may be recovered from the mixture \labelcref{eqn:hier_5}
by assuming that the ensemble size is infinite ($N = \infty$),
in which case the sample mean and covariance,
$\bx$ and $\barB$ of \cref{eqn:bx_bP_def_0},
are exact, implying a mixing distribution of Dirac delta functions:
$\delta(\bb - \bx) \delta(\B - \barB)$,
and hence the effective prior: $\NormDist (\x | \bx, \barB)$.

The EnKF-$N$ does not make this approximation,
but instead acknowledges that $N$ is finite (whence the ``finite-size'' moniker).
The mixing distribution is obtained with Gaussian sampling theory
and a non-informative hyperprior, $\pdf(\bb,\B)$.
For now, $N>M$ is assumed,
in which case $\barB^{-1}$ exists \citep[almost surely, per theorem 3.1.4 of][]{muirhead1982aspects}.

The connection to inflation comes from noting, as will be proven later,
that \cref{eqn:hier_5} reduces to:
\begin{align}
	\pdf(\x|\E)
	&=
	\int_{\alpha > 0}
	\NormDist (\x | \bx, \alpha \barB)
	\pdf (\alpha | \E)
	\diff \alpha
	\, ,
	\label{eqn:hier_14}
\end{align}
which is a mixture of candidate Gaussians over a scalar, scale parameter, only.
\begin{figure*}[htbp]
	\centering
	\includegraphics[width=1.0\textwidth]{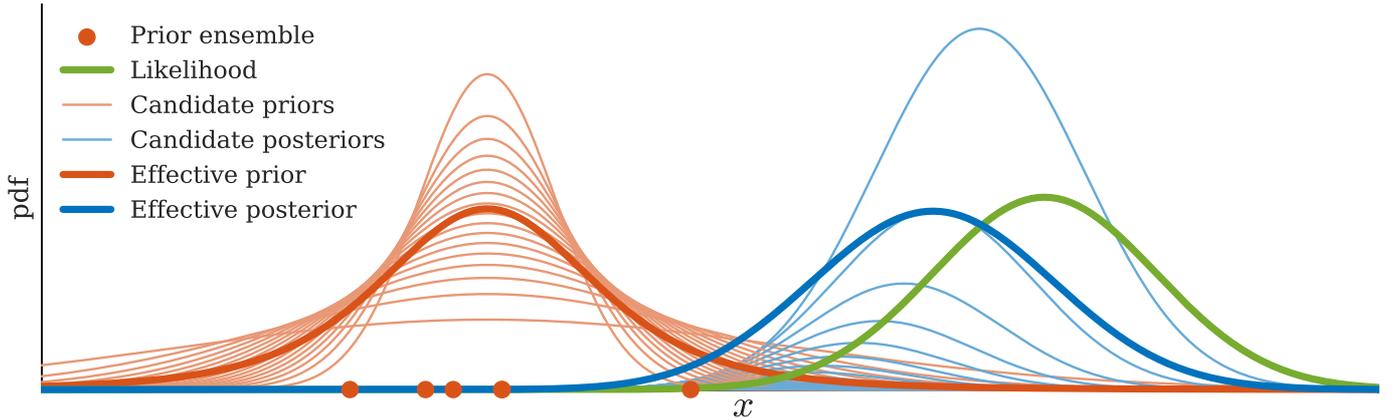}
	\caption[Illustration of the EnKF-$N$ as a scale mixture.]
	{
		Illustration of the EnKF-$N$ as a scale mixture of Gaussians, as described in \cref{sec:Sketch}.
	}
	\label{fig:post_scale_mix_illust}
\end{figure*}
The mixture \labelcref{eqn:hier_14} is illustrated by the orange objects in \cref{fig:post_scale_mix_illust}.
The candidate (prior) Gaussians are distinguished solely by the scaling,
$\alpha$, of the covariance, $\barB$.
Only a finite selection of the continuous family of candidate priors is plotted,
the selection being representative of the mixing distribution, $\pdf (\alpha | \E)$.
Interestingly, as detailed later,
this yields an effective prior, $\pdf(\x|\E)$, which is not Gaussian,
but rather a (Student's) $t$ distribution.

The effective posterior,
$\pdf(\x|\E,\y) \propto \pdf(\y|\x) \pdf(\x|\E)$,
is given by Bayes' rule, i.e. pointwise multiplication.
But the likelihood,
\begin{align}
	\pdf(\y|\x) = \NormDist(\y|\bH\x,\R) \, ,
	\label{eqn:lklhd_x}
\end{align}
per \cref{eqn:obs_eqn}, is Gaussian.
The posterior is then neither Gaussian nor $t$,
and does not simplify parametrically.
This poses a computational challenge in high-dimensional problems,
and the question of how the posterior (or an ensemble thereof)
is to be computed in practice.
Progress can be made by noting that
the averaging over the prior moments can be ``delayed'' until after application of Bayes' rule, i.e.
\begin{multline}
	\pdf(\x|\E,\y) \propto
	\int
	\underbrace{
		\NormDist \big(\y | \bH \x, \R \big)
		\NormDist (\x | \bx, \alpha \barB)
	}_{\pdf(\x,\y|\alpha,\E)}
	\pdf (\alpha | \E) \diff \alpha
	\label{eqn:prior_marg_y}
	\, .
\end{multline}
Thus, the effective posterior can also be seen as the average of the (Gaussian) candidate posteriors,
$\pdf(\x|\alpha,\y,\E)$,
each of which is given by the Kalman filter formulae for a given $\alpha$,
and computable essentially simultaneously for all $\alpha$.

The by-product of Bayes' rule is the ``evidence'', $\pdf(\y|\alpha,\E)$.
In this context, it is not a constant,
but instead constitutes the likelihood of the mixing parameter, $\alpha$.
To reflect this,
the candidate posterior curves in \cref{fig:post_scale_mix_illust}
have not been normalized to integrate to $1$,
but instead $\pdf(\y|\alpha,\E) \cdot c$.

The constant $c$ has been inserted and set such that
the particular candidate posterior whose mode coincides with that of the effective posterior also shares its height.
This makes it visible that no candidate posterior is fully coincident with the effective posterior.
Nevertheless, it seems a reasonable approximation.
But this candidate posterior corresponds to a candidate prior,
which merely amounts to choosing a particular prior inflation, $\alpha_{\star}$.
The approximation can thus be written:
\begin{align}
	\pdf(\x|\E,\y)
	&\approx
	\pdf(\x|\alpha_\star,\E,\y)
	\label{eqn:post_EB}
	\, ,
\end{align}
meaning that the integral over the hyperparameter, $\alpha$,
for the effective posterior \labelcref{eqn:prior_marg_y},
is replaced by using a particular value, $\alpha_{\star}$,
which is chosen after taking into account $\y$.
This approximation is a form of ``empirical Bayes'',
known as such because the effective prior
is approximated in a way that depends on the observations, $\y$.
This may appear to over-use the observations, $\y$,
but it is merely an artefact of the approximation.
Indeed, decomposing the integrand in \cref{eqn:prior_marg_y}
as $\pdf(\x|\alpha,\y,\E) \pdf(\y|\alpha,\E)$
makes it apparent that $\alpha$ depends on $\y$.

A posterior ensemble corresponding to the approximate posterior \labelcref{eqn:post_EB}
may be computed using standard EnKF formulae, except with $\barB$ replaced by the selected value,
$\alpha_{\star} \barB$.
Provided that the 
choice among the approximating Gaussian posteriors is judicious,
it stands to reason that the resulting ensemble yields
an improved analysis compared to that of the standard EnKF.
After all, the standard EnKF chooses its covariance estimate ($\B = \barB$) before taking into account $\y$.
By contrast, the EnKF-$N$ lets $\y$ inform this choice ($\B = \alpha_\star \barB$).
For the same reason,
even though the EnKF-$N$ does not target any particular unbiasedness,
improvement could be achieved compared to the methods targeting ``single-cycle unbiasedness'',
described below \cref{eqn:EnKF_bias_3}.

However, the main asset of the EnKF-$N$ is that its secondary dependence in $\y$
implicitly establishes a \emph{negative feedback} loop
via the sequential cycling of DA:
if the covariance estimate was too small at time $k$,
this will likely be detected and adjusted for at $k+1$.
Moreover, this feedback is ``theoretically tuned'':
parameters that may be tuned exist (cf. \cref{sec:posterior}),
but none strictly require it.

As will be shown, the inflation prior is centred on $1$,
conferring important advantages to the EnKF-$N$.
However, this anchoring to $1$ also reflects the main drawback of the EnKF-$N$:
the hyperprior is static,
so that no explicit accumulation of past information takes place for the inflation factor,
which otherwise could have been used to account for model error.
Redressing this is the subject of \cref{sec:lit_review_0} and onwards.

\subsection{The mixing distribution}
This subsection and the next further describe \cref{eqn:hier_5} for the effective \emph{prior}, $\pdf(\x|\E)$.
They are largely sourced from textbooks on Gaussian sampling theory and inference,
under the heading of ``predictive posterior'':
the probability of another draw, $\x$,
from the same distribution as the sample, $\E$
\citep[e.g., \S 3.2 of][]{gelman2003bayesian}.
The presentation is didactic, giving meaning to intermediate stages.
A concise version is provided by \citetalias{bocquet2011ensemble}.

The mixing distribution in \cref{eqn:hier_5} is given by:
\begin{align}
	\pdf (\bb, \B | \E) \propto \pdf(\E|\bb,\B) \pdf(\bb,\B)
	\label{eqn:mix_1}
	\, ,
\end{align}
where $\pdf(\bb,\B)$ is a hyperprior to be specified.
Here, as in \citetalias{bocquet2011ensemble},
the Jeffreys priors are independently assigned to the hyperparameters:
\begin{align}
	\pdf (\bb, \B)
	=
	\pdf (\bb) \pdf (\B)
	\propto
	1 \cdot \abs{\B}^{-(M+1)/2}
	\, .
	\label{eqn:hyperp}
\end{align}
This is a prior designed to be as non-informative (agnostic) as possible. 
It may be derived by positing invariance in location and scale
\citep[e.g., \S 12.4 of][]{jaynes2003probability}.
\citetalias{bocquet2015expanding} also showed the utility of
using a highly informative hyperprior, suitable in contexts with little nonlinearity.
Examples were also given for encoding information such as
climatology or conditional statistics,
resulting in a form of localization.

By the Gaussian ensemble assumption \labelcref{eqn:const_4},
\begin{align}
	\pdf(\E|\bb,\B) &= \prod_n \NormDist(\x_n|\bb,\B)
	\notag
	\\
	&\propto \abs{\B}^{-N/2} \, e^{-\sum_n \norm{\x_n - \bb}^2_\B /2}
	\label{eqn:fapwrg}
	\, ,
\end{align}
where $\norm{\x}_\B^2 = \x\tr\B^{-1}\x = \trace(\x\x\tr\B^{-1})$.
Now, writing $\x_n - \bb = ( \bx - \bb ) + ( \x_n - \bx )$,
it can be shown that
\begin{align}
	\sum_n \norm{\x_n - \bb}^2_{\B}
	&=
	N \norm{\bx - \bb}^2_{\B} + \trace(\cN \barB \B^{-1})
	\, .
	\label{eqn:xn_id2}
\end{align}
Combining \cref{eqn:hyperp,eqn:fapwrg,eqn:xn_id2}
for the mixing distribution \labelcref{eqn:mix_1},
the resulting factors may be identified as:
\begin{align}
	\pdf (\bb, \B | \E)
	&=
	\underbrace{
		\NormDist (\bb | \bx, \B/N)
	}_{\pdf (\bb | \B, \E)}
	\underbrace{
		\InvWish (\B | \barB,  \compactN)
	}_{\pdf (\B | \E)}
	\label{eqn:awbuhawef}
	\, ,
\end{align}
where $\InvWish$ is the inverse-Wishart distribution (cf. \cref{tab:pdfs} of \cref{sec:standard_pdfs}).

\subsection{Integrating over the mean}
Writing the integrand of \cref{eqn:hier_5} as
$\pdf(\x,\bb,\B|\E) = \pdf (\bb | \x, \B, \E) \pdf (\x | \B, \E) \pdf (\B | \E)$,
the integral over $\bb$ becomes trivial, leaving just the latter two factors:
\begin{align}
	\pdf(\x|\E)
	&=
	\int
	\pdf (\x | \B, \E)
	\pdf (\B | \E)
	\diff \B
	\, ,
	\label{eqn:hawg32}
\end{align}
of which $\pdf (\B | \E)$ was obtained in \cref{eqn:awbuhawef}.

Meanwhile, recalling $\pdf (\x | \bb, \B)$ 
and $\pdf (\bb | \B, \E)$ from \cref{eqn:const_4,eqn:awbuhawef} respectively,
it may be shown by completing the square in $\bb$ that
\begin{multline}
	\pdf (\x | \bb, \B)
	\pdf (\bb | \B, \E)
	=\\
	\underbrace{
		\NormDist \big(\bb \big| {\textstyle \frac{N\bx + \x}{N+1}}, \B/(N+1)\big)
	}_{\pdf (\bb | \x, \B, \E)}
	\underbrace{
		\NormDist (\x | \bx, \eN \B)
	}_{\pdf (\x | \B, \E)}
	\, ,
\end{multline}
where $\eN = 1 + \frac{1}{N}$.
The underbraces follow by identification
and provide the first factor in \cref{eqn:hawg32}.

Thus,
\begin{align}
	\pdf(\x|\E)
	&=
	\int
	\NormDist (\x | \bx, \eN \B)
	\InvWish (\B | \barB,  \compactN)
	\diff \B
	\, .
	\label{eqn:h50}
\end{align}
It should be appreciated that 
\cref{eqn:h50} would be unchanged
if $\bb = \bx$ had been assumed from the start,
except for the slight adjustment of $\eN$
and the reduction from $N$ to $\compactN$ in the ``certainty'' parameter of $\InvWish$.
By contrast, as shown in the following,
the uncertainty in $\B$ has significantly more interesting consequences.

\subsection{Reduction to a scale mixture}
\label{sec:Prior_as_scale}

This section derives the scale mixture \cref{eqn:hier_14}.

While conventional,
the assumption ``$\barB \propto \B$'' is ill-suited for inflation targeting sampling error,
as it yields an inflation prior with an overpowering confidence,
to the detriment of the likelihood \citep[][\S C.4]{raanes2016thesis}.
This assumption is therefore not made.
But then merely defining the inflation parameter becomes challenging.
Clearly, it must be some scalar summary statistic on the ``ratio'' of $\B$ versus $\barB$;
possibilities include using the determinant, trace, or matrix norms.
However, the subsequent \emph{assignment} ``$\B = \scl \barB$'' would represent an artificial approximation.
By contrast, the following definition and developments make no approximations.

Consider a fixed $\x$,
and define the (squared) inflation:
\begin{align}
	\scl = \frac{\norm{\x - \bx}^2_{\barB}}{\norm{\x - \bx}^2_{\B}}
	\, .
	\label{eqn:lambda_def_2}
\end{align}
%
Now, given the ensemble, $\E$, the sample moments
$\bx$ and $\barB$ are known (fixed),
while $\pdf (\B | \E) = \InvWish (\B | \barB,  \compactN)$
per \cref{eqn:awbuhawef}.
Thus, by the reciprocity of the Wishart distribution
(\cref{prop:Wish} of \cref{sec:standard_pdfs}),
$\B^{-1} \sim \Wishart(\barB^{-1},\compactN)$.
\Cref{prop:WishChi2} can then be applied to yield
$1/\scl \sim \CS(1, \compactN)$.
Thus, again by reciprocity (\cref{prop:CS}),
\begin{align}
	\pdf(\scl|\E)
	&=
	\InvCS(\scl | 1, \compactN)
	\label{eqn:lam3}
	\, ,
\end{align}
meaning that $\scl$ is inverse-chi-square (cf. \cref{tab:pdfs}),
with location parameter $1$ and certainty $N-1$.
%
\newcommand{\JacB}{\mathop{}\! J}

But the \PDF{} $\pdf(\scl|\E)$
could also have been derived
by marginalizing $\pdf(\B|\E)$ over $\C \in \RestSpace_\scl$,
where $\C$ denotes (any parameterization of)
the degrees of freedom in $\B$ not fixed by $\scl$,
i.e.
$\RestSpace_\scl = \{\C \in \Reals^{M(M+1)/2-1} \; ; \;
\B(\scl,\C) \in \CovSpace, \; \norm{\x-\bx}^2_{\B} = \norm{\x-\bx}^2_{\scl \barB}\}$.
Formally,
\begin{align}
	\int_{\RestSpace_\scl}
	\pdf(\B|\E)
	\JacB
	\diff \C
	=
	\pdf(\scl|\E)
	\, ,
\end{align}
with $\JacB$ denoting the Jacobian determinant of $(\scl,\C) \mapsto \B$.
Inserting the \PDF{}s from \cref{eqn:awbuhawef,eqn:lam3}:
\begin{align}
	\int_{\RestSpace_\scl}
	\InvWish(\B|\barB, N{-}1)
	\JacB
	\diff \C
	\label{eqn:pdf_s_1_2}
	&=
	\InvCS(\scl|1,N{-}1)
	\, .
\end{align}
Now, the covariance mixture \labelcref{eqn:h50} can be rearranged as:
\begin{multline}
	\pdf(\x|\E)
	\propto\\
	\int_\CovSpace
	\exp\big({\mahalf \norm{\x-\bx}^2_{\eN \B}}\big)
	\InvWish\big(\B \big| \textstyle \frac{N-1}{N} \barB,  N \big)
	\diff \B
	\, .
	\notag
\end{multline}
The same change of variables
then yields:
\begin{multline}
	\pdf(\x|\E)
	\propto
	\int_{\scl>0}
	\exp\big({\mahalf \norm{\x-\bx}^2_{\eN \scl \barB}}\big)
	\\
	\left(
	\int_{\RestSpace_\scl}
	\InvWish\big(\B \big| \textstyle \frac{N-1}{N} \barB,  N \big)
	\JacB
	\diff \C
	\right)
	\diff \scl
	\, .
	\label{eqn:after_CVar_2}
\end{multline}
The inner integral can be substituted by comparing it to 
\cref{eqn:pdf_s_1_2}, yielding:
\begin{align}
	\pdf&(\x|\E) \notag \\
	&\propto
	\int
	\exp\big({\mahalf \norm{\x-\bx}^2_{\eN \scl \barB}}\big)
	\InvCS\big(\scl \big| \textstyle \frac{N-1}{N}, N \big)
	\diff \scl
	\notag
	\\
	&\propto
	\int
	\NormDist (\norm{\x - \bx}_{\barB} \,|\,  0, \eN \scl)
	\InvCS(\scl | 1, \compactN)
	\diff \scl
	\, .
	\label{eqn:h61}
\end{align}
In conclusion, the covariance mixture of \cref{eqn:h50} reduces to a scale mixture.
%
An alternative, direct proof,
using tricks from complex analysis instead of \cref{prop:WishChi2},
was given in a preprint version of this paper.

Note that the scale mixture \labelcref{eqn:h61} has been written
using the notational trick
where $\NormDist$ acts as a \emph{univariate} function.
Also, since $\scl$ is defined via $\x$,
the integrand of \cref{eqn:h61} cannot be read as
``$\pdf(\x|\E,\scl)\pdf(\scl|\E)$''.
%
By contrast,
the mixture \labelcref{eqn:hier_14} is obtained by undoing the trick, and defining $\alpha = \eN \scl$.


\subsection{Ensemble subspace parameterization}
\label{sec:CVar}
Let $\ones$ be the vector of ones of length $N$, and $\I_N$ the $N {\times} N$ identity matrix.
Then the sample moments,
given in \cref{eqn:bx_bP_def_0}, may be conveniently expressed as:
\begin{align}
	\label{eqn:A_def_2}
	\bx   &= \E \ones/N\, ,&
	\barB &= \fracN \X \X\tr \, ,
\end{align}
where
	$\X = \begin{bmatrix}
		\x_1 -\bx, & \ldots & \x_n -\bx, & \ldots & \x_N -\bx
	\end{bmatrix} = \E \bPi_\ones^\perp$
is the ensemble ``anomalies'', 
with $\bPi_\ones^\perp = \AN$
the orthogonal projector\label{symbl:proj_mat}
onto $\range(\ones)^\perp$, the orthogonal complement space to $\range(\ones)$.

So far it has been assumed that $N>M$
so that $\barB$ is invertible (almost surely) and that the ensemble spans the entire state space.
This is unrealistic for geoscientific DA,
where $N$ rarely exceeds 100, while $M$ may exceed $10^9$.
More reasonably, it is henceforth assumed
that the support of the forecast \PDF{} is confined to the ensemble subspace,
i.e. the affine space $\{\x \in \Reals^M \tq [\x - \bx] \in \range(\X)\}$.
This assumption is actually conventional,
as it is implied by the standard EnKF's assumption that $\bb = \bx$ and $\B = \barB$ along with Gaussianity.
The assumption means that the ensemble has sufficient rank.
Thus, one may expect tolerable accuracy of the filter,
even without localization \citep{bocquet2016aus4d}.
%
It is preferable to work with variables that embody the restriction of the assumption
\citep{hunt2007efficient};
therefore,
with $\w \in \Reals^N$,
the following change of variables is done:
\begin{align}
	\x(\w) = \bx + \X \w
	\, .
	\label{eqn:CVar_w}
\end{align}

In terms of the new variable,
the likelihood \labelcref{eqn:lklhd_x} may be succinctly written as:
\begin{align}
	\pdf(\y|\w) = \NormDist (\bbdelta | \Y \w, \R) \, .
	\label{eqn:lklhd_w}
\end{align}
with $\bbdelta = \y - \obsmf{\E}\ones/N$ the average innovation,
and $\Y = \obsmf{\E} \PiAN = \obsm{} \X$ the corresponding observation anomalies.

For the effective prior \labelcref{eqn:h61},
note that the ensemble members expressed in the coordinate system of $\w$ are merely
the coordinate vectors ($\x_n = \bx + \X \e_n$, with $\e_n$ being the $n$-th column of $\I_N$).
Hence, in this coordinate system,
the sample mean is $\ones/N$, replaceable by zero since $\X\ones=\bvec{0}$,
and the sample covariance matrix is $\fracN \I_N$.
Substituting these for $\bx$ and $\barB$ in \cref{eqn:h61}
is a shortcut to obtain the effective prior for $\w$;
with $\alpha = \eN \scl$,
\begin{align}
	\pdf&(\w|\E)
	\propto
	\label{eqn:h71}
	\\
	&\int
	\alpha^{-g/2}
	\NormDist \big(\|{\w}\|_{\frac{1}{N-1} \I_N} \,\big|\,  0, \alpha \big)
	\InvCS(\alpha | \eN, \compactN)
	\diff \alpha
	\, ,
	\notag
\end{align}
where the presence of $\alpha^{-g/2}$ is explained in the following.

Denote $g$ the dimensionality of the nullspace of $\X$.
Due to $\PiAN$ it holds that $g = \max(1, N{-}M)$, almost surely.
Thus, typically $g=1$, and the parameterization in $\w$ has one direction of redundancy,
warranting careful attention.
The issue is analogous to expressing 1 random variable as the sum of 2,
or indeed expressing $N-g$ random variables as a linear combination of $N$.
The principle is that regardless of how the probability space is augmented with the redundant degrees of freedom,
once these are marginalized out, one should be left with the original distribution.
\citetalias{bocquet2015expanding} showed that the adjustment of $g$ in \cref{eqn:h71} is then required.

\subsection{The saddlepoint form}
\label{sec:saddle}

\newcommand*{\ff}{\mathop{}\! f}
\newcommand*{\Ff}{\mathop{}\! F}
\newcommand*{\zf}{\mathop{}\! \zeta}
\newcommand*{\af}{\mathop{}\! \alpha}


Denote $f$ the integrand
of the scale mixture \labelcref{eqn:h71}.
Expanding the parametric \PDF{}s yields:
\begin{align}
	\ff(\alpha,z)
	&=
	\alpha^{-(N+g)/2-1}
	\, e^{ - z / 2 \alpha }
	\, ,
	\label{eqn:f_integrand}
\end{align}
where all of the dependency in $\w$ is contained in
\begin{align}
	z(\w) &= \cN \big( \eN + \|\w\|^2 \big)
	\, .
\end{align}
Defining $F(z(\w)) \propto \pdf(\w|\E)$ for the effective prior,
\cref{eqn:h71} may be restated as:
\begin{align}
	\Ff(z) =
	\int
	\ff(\alpha,z)
	\diff \alpha
	\label{eqn:sm_l1}
	\, .
\end{align}
It can be seen that the change of variables $u = \alpha/z$
factors $z$ out of the integral,
yielding
\begin{align}
	\Ff(z)
	&=
	z^{-(N+g)/2}
	\Ff(1)
	\, ,
	\label{eqn:pw_zF}
\end{align}
or, reverting to the full $\w$ notation,
\begin{align}
	\pdf(\w|\E)
	&\propto
	\big(
		\eN + \|{\w}\|^2
	\big)^{-(N+g)/2}
	\, ,
	\label{eqn:effective_prior_w}
\end{align}
which is a $t$ distribution
(cf. \cref{sec:standard_pdfs}),
also called a Cauchy distribution when $g=1$.
The $t$ distribution is elliptical, like the Gaussian
\citep[][\S 1.5]{muirhead1982aspects},
but has thick tails,
making it suited for robust inference
\citep{geweke1993bayesian,fernandez1999multivariate,roth2017robust}.


Unlike \citetalias{bocquet2011ensemble},
here the $t$ distribution form \labelcref{eqn:effective_prior_w} of the effective prior
will not be used directly.
Instead, the effective prior ($F$) will again be expressed
as a Gaussian ($G$) with inflation.
To that end, note that,
for general functions $F$ and $G$ with $\image[F] \subseteq \image[G]$,
there will always exist a function $\zf(z)$ such that
$F(z) = G(\zf(z))$.

To find a suitable $G$,
consider applying the mean-value theorem to the integral \labelcref{eqn:sm_l1} for a fixed $z$,
denoting $\zeta^{-1}$ the particular point for $\alpha$ .
This will not work because the integrational interval,
$[0,\, \infty)$, is of infinite length.
In place of the length, therefore,
substitute $c_1 \zf^{-1}$ to form:
$
	G (\zeta, z)
	=
	c_1 \zeta^{-1} \ff(\zeta^{-1},z)$,
where the constant $c_1$
ensures that the height (and hence image) of $G( \,.\, ,z)$ is sufficient.
Inserting $\ff$ from \cref{eqn:f_integrand} yields
\begin{align}
	G(\zeta,z)
	&=
	c_1
	\zeta^{(N+g)/2}
		\, e^{
			- z \zeta / 2
		}
		\, ,
		\label{eqn:Gzz}
\end{align}
This $G$ works well;
indeed, equating \cref{eqn:Gzz} to \labelcref{eqn:pw_zF}
immediately yields the associated function $\zf(z) = c_2/z$.

In summary, the effective prior
may be expressed by $G$ \labelcref{eqn:Gzz} which,
similarly to the integrand, $f$,
is Gaussian in $\w$.
For later optimization purposes,
$c_2=N{+}g$ is set,
and the logarithm is taken:
\begin{subequations}
	\begin{align}
		\pdf(\w|\E)
		&= 
		c_1
		\exp\big(\mahalf L \big)
		\, ,
		\label[equation]{eqn:Lap_aprx_0}
		\\
		L(\zeta,\w)
		&=
		\big( \eN + \|{\w}\|^2 \big) \zeta
		- (N+g) \log \zeta
		\, ,
		\label[equation]{eqn:g_fawef}
		\\
		\zeta(\w)
		&=
		\frac{N+g}{\eN + \|{\w}\|^2}
		\label[equation]{eqn:zeta_w_0}
		\, ,
	\end{align}
	\label[subeqns]{eqn:lap_aprox_all}%
\end{subequations}
The value of $c_2$ was chosen so as to yield the property that
$\pd{L}{\zeta}(\zeta(\w),\w) = 0$ for any $\w$,
as can be directly verified.
Conversely, this means that $\zeta$ may be treated as a free variable to be optimized for,
because \cref{eqn:zeta_w_0} is satisfied wherever $\pd{L}{\zeta} = 0$.
This tactic becomes useful in the following section.

The above form of the effective prior
may be derived (as in a preprint version of this paper)
as a ``saddlepoint approximation''.
Here, however, there is no approximation.
Its exactitude is a remarkable feature
known to arise in a few cases \citep{azevedo1994laplace,goutis1999explaining}.

\subsection{The posterior and its mode}
\label{sec:posterior}


Define $J = -2 \log \pdf(\w|\E,\y)$ plus a constant,
where the posterior is given by Bayes' rule
with the likelihood \labelcref{eqn:lklhd_w}
and the effective prior \labelcref{eqn:lap_aprox_all}.
The log posterior reads:
\begin{subequations}
	\begin{align}
		&J(\zeta,\w)
		=
		L(\zeta,\w)
		-2 \log \NormDist (\bbdelta | \Y \w, \R)
		+ c
		\label[equation]{eqn:post_4}
		\\
		&=
		\eN \zeta - (N+g) \log \zeta
		+ \zeta \|\w\|^2
		+	\norm{\bbdelta - \Y \w}_{\R}^2
		\label[equation]{eqn:post_7}
		\, .
	\end{align}
	\label[subeqns]{eqn:post_47}%
\end{subequations}
Completing the square in $\w$ yields:
\begin{align}
	J(\zeta,\w)
	&= \norm{\w - \bw^\tn{a}(\zeta)}^2_{\barPa_\w(\zeta)} + D(\zeta)
	\label{eqn:post_15}
	\, ,
\end{align}
where the quadratic form is specified
by the usual EnKF subspace analysis formulae:
\begin{subequations}
	\begin{align}
		\barPa_\w(\zeta) &= \left( \zeta \I_N + \Y\tr \R^{-1} \Y \right)^{-1}
		\label[equation]{eqn:EnKF-N-w-KF-Cov}
		\, , \\
		\bw^\tn{a}(\zeta) &= \barPa_\w(\zeta) \, \Y\tr \R^{-1} \bbdelta
		\, ,
		\label[equation]{eqn:EnKF-N-w-KF-Mu}
	\end{align}
	\label[subeqns]{eqn:EnKF_subspace}%
\end{subequations}
and $D$ should be recognized as
the ``dual'', as in \citet{bocquet2012combining}:
\begin{align}
	D(\zeta)
	&= \eN \zeta - (N+g) \log \zeta
	+ \norm{\bbdelta}^2_{\R + \Y \Y\tr/\zeta}
	\, .
	\label{eqn:Dual_def}
\end{align}

Now, $\zeta$ depends on $\w$,
and so the maximization of $\pdf(\w|\E,\y)$ is not as obvious as \cref{eqn:post_15} suggests.
Fortunately, to find a critical point, it suffices to satisfy
\begin{subequations}
	\begin{align}
		\pd{J}{\w}    &= \bvec{0} \label[equation]{eqn:Jw0} \, , \\
		\pd{J}{\zeta} &= 0 \label[equation]{eqn:Jz0} \, . 
	\end{align}
	\label[subeqns]{eqn:J0}%
\end{subequations}
This is because
the criteria \labelcref{eqn:Jw0,eqn:Jz0}
imply $\dd{J}{\w} = \pd{J}{\w} + \pd{J}{\zeta}\dd{\zeta}{\w} = \bvec{0}$,
where $\zeta(\w)$ is given by \cref{eqn:zeta_w_0},
which is enforced since $\pd{L}{\zeta}=0$,
as follows from \cref{eqn:post_4,eqn:Jz0}.

Now, the first criterion \labelcref{eqn:Jw0}
is trivially satisfied by setting
$\w = \bw^\tn{a}(\zeta)$ for a given $\zeta$, as seen from \cref{eqn:post_15}.
But it can also be seen that $J = D$ along the constraint $\w = \bw^\tn{a}(\zeta)$, and so 
\begin{align}
	\dd{D}{\zeta} = \dd{J}{\zeta} = \pd{J}{\zeta} + \pd{J}{\w} \dd{\w}{\zeta} = \pd{J}{\zeta}
	\, .
\end{align}
Hence, finding $\zeta$ such that $\dd{D}{\zeta} = 0$ will satisfy the second criterion \labelcref{eqn:Jz0}.

In conclusion, $\w = \bw^\tn{a}(\zeta_\star)$
is a critical point of the effective posterior
if and only if $\zeta_\star$ is a local minimizer of $D(\zeta)$.
Since both of the terms $D(\zeta)$ and $\norm{\w - \bw^\tn{a}}^2_{\barPa_\w}$ of
\cref{eqn:post_15} are here individually minimized,
this critical point
must be a minimum,
as was originally shown using Lagrangian duality theory by \citet{bocquet2012combining}.
Hence the $N$-dimensional, non-Gaussian mode-finding problem for $\pdf(\w|\E,\y)$
may be exchanged for the scalar optimization problem in $\zeta$.

The optimization of $D(\zeta)$ requires iterating,
but each evaluation of \labelcref{eqn:Dual_def} and its derivative
is computationally negligible,
given the singular value decomposition (SVD),
\begin{align}
	\U \diag(\bar{\sigma}_1,\ldots,\bar{\sigma}_P) \V\tr = [\cN\R]\msq \Y
	\label{eqn:svd_nRY}
	\, ,
\end{align}
has been obtained beforehand,
as is typical to compute \cref{eqn:EnKF_subspace}.
Multiple minima are a rarity;
in such cases $\zeta_\star$ will depend on the optimizer and initial guess,
here Newton's method and $\zeta = N-1$.

To obtain an analysis posterior \emph{ensemble},
a Gaussian approximation to the effective posterior is chosen.
In addition to its simplicity,
twin experiments \citepalias{bocquet2011ensemble,bocquet2012combining,bocquet2015expanding}
have provided solid support to using that of $\zeta_\star$:
\begin{align}
	\pdf(\w|\E,\y) \approx \NormDist\big(\w \big| \bw^\tn{a}(\zeta_\star),\barPa_\w(\zeta_\star)\big)
	\, .
	\label{eqn:post_enkfn}
\end{align}
Notably, this approximation matches the mode and Hessian
of the exact $\pdf(\w|\E,\y)$.
The corresponding ensemble is constructed as:
\begin{align}
	\E^\tn{a}
	=
	\left[\bx+\X \bw^\tn{a}(\zeta_\star)\right] \ones\tr
	+ \sqrt{\compactN}\, \X \T 
	\, ,
	\label{eqn:Ea_sqrt_zeta}
\end{align}
\noindent 
where $\T = (\barPa_\w(\zeta_\star))\sq$
is readily computed using the same SVD \labelcref{eqn:svd_nRY} as before,
and may be appended by a mean-preserving orthogonal matrix \citep{sakov2008implications}.
Note that this is ``just'' the symmetric square-root EnKF,
i.e. the ensemble transform Kalman filter (ETKF) of
\citet[][]{bishop2001adaptive,hunt2004four},
except for a prior (squared) inflation of
\begin{align}
	\alpha_\star = \cN/\zeta_\star \, .
\end{align}

As discussed by \citetalias{bocquet2015expanding},
the choice of candidate posterior is not unassailable,
and certain modifications of the dual function can be argued
on the grounds of modifying this choice.
Indeed, if the influence of the likelihood,
quantified by
\begin{align}
	\bar{\sigma}^2 = \trace(\bH \barB \bH\tr \R^{-1})/P \, ,
	\label{eqn:bsig_def}
\end{align}
and computed using the SVD \labelcref{eqn:svd_nRY},
is small:  $\bar{\sigma}^2 \rightarrow 0$,
then it becomes crucial to adjust the choice of inflation factor towards $1$.
Moreover, weakly nonlinear contexts create relatively little sampling error.
In such cases, the hyperprior may be too agnostic.
This can be corrected for by increasing (tuning)
the certainty of $\InvCS$ to a higher value than $\compactN$,
yielding a Dirac-delta in the limit.
Finally,
since $\zeta$ is not actually constant in $\w$,
the Hessian, $(\barPa_\w(\zeta_\star))^{-1}$,
should be corrected by subtracting $e \bw^\tn{a}{{}\bw^\tn{a}}\tr$,
where $e = \frac{2\zeta_\star^2}{N+g}$,
\citepalias{bocquet2015expanding}.
Since this is but a rank-one update,
a corrected transformation can be
computed without significant expense as
$\T (\I_N + \gamma \bv \bv\tr)$,
where $\bv = \T\tr \bw^\tn{a}$, $\gamma = e/(\tau^{1/2} + \tau)$,
and $\tau = 1 - e \bv\tr\bv$,
as may be deduced from equation (B2) of \citet{bocquet2016localization}.
However, this is typically a very minor correction,
while the Gaussian posterior remains but an approximation.
None of these adjustments were deemed necessary to use in the numerical experiments of this paper.

\section{Overview: adaptive inflation}
\label{sec:lit_review_0}

The filtering problem as formulated with \cref{eqn:state_sde_discrete,eqn:obs_eqn}
uses additive noise processes $\bxi_k$ and $\up_k$
to represent model and observation errors.
``Primary'' filters \citep{dee1985efficient} assume the noise/error covariances, $\Q$ and $\R$,
are perfectly known.
In reality, these system statistics are rarely well known;
this deteriorates the state estimates and can cause filter divergence.
Self-diagnosing and correcting filters
which estimate and modify the given noise statistics are called ``adaptive''.
Adaptive Kalman filter literature include \citet{mehra1970approaches,mohamed1999adaptive}.
Adaptive particle filter literature include \citet{storvik2002particle,ozkan2013marginalized}.

This section is concerned with adaptive filtering for DA,
focusing on the EnKF and the on-line\footnote{%
	``On-line'' means that the estimation is included in the
	DA cycling loop, so as to be updated each time new data $\y$ is received.
	Some off-line approaches not further reviewed include
	\citet{dee1999maximum1,ueno2010maximum,mitchell2014accounting}.
}
estimation of the model error, i.e. $\Q$.\footnote{%
The methods may also possess some skill in dealing with errors
due to non-Gaussianity, biases, and even misspecification of $\R$.
}
Among the literature using full
\citep[][]{miyoshi2013estimating,nakabayashi2017extension,pulido2018stochastic}
and/or diagonal
\citep{ueno2016bayesian,dreano2017estimating} covariance parameterizations,
some success has been noted.
However, the scope of this paper is restricted to the estimation of a single
multiplicative inflation factor, $\beta$, for $\barB$.
This assigns a structure to the covariance and reduces the dimensionality and complexity
of the problem, thus regularizing it.
The tradeoff is a bias,
but this drawback is largely offset by spatialization.

Since this section presents adaptive inflation for model error,
the inflation factor is here labelled $\beta$.
This contrasts it to $\alpha$ of the EnKF-$N$, which targets sampling error.
Formally, the unknown, $\beta$, is defined by the modelling assumption
that the ensemble is now drawn with a covariance
that is $\beta$ times too small,
\begin{align}
	\x_n \sim \NormDist(\bb, \B/\beta) \, ,
	\label{eqn:fawefhawe}
\end{align}
as compared to the distribution \labelcref{eqn:const_4} of the truth.
Moreover, in order to neglect sampling errors,
the ensemble is assumed infinite ($N=\infty$).
Again, this is commonly tacitly assumed in the adaptive EnKF literature.
The reason, as discussed below \cref{eqn:hier_5},
is that it yields a prior that is
(the limit of the $t$ distribution which is)
Gaussian.
In summary,\footnote{%
For familiarity,
the state distributions are written in terms of the original variable $\x$,
rather than the subspace variable $\w$ of \cref{eqn:CVar_w}.
Note that most of the following inflation estimators
also implicitly address the rank deficiency issue,
as they do not ignore components of the innovation outside of the (observed) ensemble subspace.
This may not be generally beneficial (cf. \S5 of \cref{sec:catalogue}).
}
\begin{align}
	\pdf(\x|\beta, \E)
	&=
	\NormDist\big(\x \big| \bx, \beta \barB \big)
	\, .
	\label{eqn:prior_x_beta}
\end{align}

The $N=\infty$ assumption is rolled back in the bias study of \cref{sec:Single_cycle_bias}.
More pragmatically, \cref{sec:hybrid} makes a hybrid of the EnKF-$N$ inflation
with a method of adaptive inflation for model error.
The following review and analyses serve to choose a method with which to make this hybrid.

The review is split between methods that may be termed
``marginal'', working with $\beta$ separately from $\x$,
and ``joint'', working with $(\beta,\x)$ simultaneously.
The joint approach, including ``variational'' and ``hierarchical'' methods,
is theoretically appealing.
However, on closer inspection, including numerical testing,
it was found to be less advantageous.
Therefore, the marginal methods take centre stage,
while the review of the joint methods has been placed
in \cref{sec:Joint_xs_estim}.

\subsection{Survey: marginal estimation}
\label{sec:Marginal_s_estim}
Recall the average innovation,
$\bbdelta = \y - \bH \bx$.
For brevity, the explicit conditioning on the ensemble, $\E$, is henceforth dropped.

\subsubsection{Taking the trace of covariances}
Writing $\bbdelta = (\y - \bH \x) + \bH(\x - \bb) + \bH(\bb - \bx)$
it can be seen that
\begin{align}
	\Expect [\bbdelta \bbdelta\tr | \B ] = \R + \bH \eN \B \bH\tr
	\label{eqn:fawegnbawef}
	\, .
\end{align}
A departure from non-ensemble works \citep[e.g.,][]{daley1992estimating}
is the adjustment $\eN = 1 + 1/N$,
resulting from (necessarily) using
the prior ensemble mean, $\bx$,
rather than the true prior mean, $\bb$,
to define the innovation, $\bbdelta$.
However, since $N=\infty$ is here assumed, $\eN = 1$.

Substitute the expectation $\Expect [\bbdelta \bbdelta\tr | \B ]$
by the observed value $\bbdelta \bbdelta\tr$ in \cref{eqn:fawegnbawef},
and $\B$ by $\beta \barB$, per \cref{eqn:prior_x_beta}.
Then,
\begin{align}
	\bbdelta \bbdelta\tr &\approx
	\beta \bH \barB \bH\tr + \R
	\eqqcolon \barC(\beta)
	\label{eqn:expected_innov_3}
	\, ,
\end{align}
which suggests matching (some univariate summary of) $\bbdelta \bbdelta\tr$
by adjusting the inflation, $\beta$.
For example, \citet{li2009simultaneous}
working in the framework of the ETKF,
use the estimator:
\begin{align}
	\hat{\beta}_\I
	=
	\frac{ \norm{\bbdelta}^2 - \trace(\R)} { \trace(\bH \barB \bH\tr) }
	\, ,
\end{align}
which follows directly from the trace of \cref{eqn:expected_innov_3}.
Alternatively, the estimator of \citet{wang2003comparison}
can be obtained by transforming \cref{eqn:expected_innov_3} by $\R^{-1}$ before taking the trace,
thus allowing for heterogeneous observations and different units. This yields:
\begin{align}
	\hat{\beta}_{\R} =
	\frac{\norm {\bbdelta}^2_\R/P - 1}
	{\bar{\sigma}^2}
	\, ,
	\label{eqn:s_miyoshi11}
\end{align}
where $\trace(\bbdelta \bbdelta\tr \R^{-1}) = \norm{\bbdelta}_\R^2$
has been used and $\bar{\sigma}^2$ is defined in \cref{eqn:bsig_def}.

\citet{miyoshi2011gaussian} 
proposes a variant using the Schur product $\circ \R^{-1}$.
However, the regular algebra behind $\hat{\beta}_{\R}$ is preferable,
as it decorrelates the diagonals of $\bbdelta \bbdelta\tr$
and hence diminishes the variance of their trace.
More variants are studied in \cref{sec:Other_variants}.
However, \cref{sec:Single_cycle_bias} show them
to have more bias than $\hat{\beta}_{\R}$
because of further exposure to the uncertainty in $\barB$,
which is present when the $N=\infty$ assumption is not made.

\subsubsection{Maximum likelihood estimation}
\label{sec:Maximum_likelihood_results}
As an alternative to the trace-based estimators,
consider some results of likelihood maximization.
Recalling \cref{eqn:lklhd_x,eqn:prior_x_beta},
it can be shown that the likelihood for $\beta$ is:
\begin{align}
	\pdf(\y|\beta)
	&=
	\NormDist\big(\y \,\big|\, \bH \bx, \C \big)
	=
	\NormDist\big(\bbdelta \,\big|\, \bvec{0}, \C \big)
	\, ,
	\label{eqn:mult_norm_lklhd}
\end{align}
where $\C = \barC(\beta)$ of \cref{eqn:expected_innov_3}.
By this Gaussianity, 
$\bbdelta \bbdelta\tr$ 
can be seen as the maximum of the likelihood for $\C$,
but only in the case of univariate observations ($P=1$).
In the multivariate case, the maximization is only defined
restricted to the direction of $\bbdelta \bbdelta\tr$.

Less artificially,
the maximization can be rendered well-posed
by restricting $\C$ to the scaling: $\C(\theta) = \theta \C_0$ for some $\C_0$,
in which case the maximum likelihood (ML) estimate of $\theta$ is $\norm{\bdelta}^2_{\C_0}/P$,
as noted by \citet{dee1995line}.
Unfortunately, with the parameterization
$\C(\beta) = \barC(\beta)$ of \cref{eqn:mult_norm_lklhd},
the ML estimate, $\hat{\beta}_\tn{ML}$, is not analytically available,
and will require iterations
\citep[e.g.,][]{mitchell2000adaptive,zheng2009adaptive,liang2012maximum}.

Also note that \cref{eqn:expected_innov_3} may be derived in the variational framework,
where it is seen as a consistency criterion on the cost function
\citep{desroziers2001diagnosis,chapnik2006diagnosis,menard2016error}.
These references are also known for deriving further diagnostics,
employing distances labelled ``observation-analysis'' and ``analysis-background'',
which enable the simultaneous estimation of the observation error covariance.
\citet{li2009simultaneous} make use of this in an ensemble framework,
as do \citet{ying2015adaptive,kotsuki2017adaptive},
who makes use of the scheme in a relaxation variety.

\subsubsection{Secondary filters}
The accumulation and memorization of past information has gradually become more sophisticated.
\citet{wang2003comparison} take the geometric average of $\hat{\beta}_{\R}$ over time,
while \citet{mitchell2000adaptive} use the median of instantaneous maximum likelihood estimates.
Lacking the fuller Bayesian setting, neither yields consistency (convergence to the true value) in time,
because a temporal average of some point estimate $\hat{\beta}_k$ based on $\pdf(\y_k|\beta)$
is generally not the value to which $\beta|\y_{1:K}$ converges as $K \rightarrow \infty$.
Nevertheless, this mismatch is not likely to be severe,
and the simplicity of the approach is an advantage.

The approach of temporally averaging likelihood estimates has since
been replaced by the more rigorous approach of filtering.
It should be noted that this ``secondary'' filter is valid
because the innovations are supposedly independent in time \citep{mehra1970approaches}.
\citet{li2009simultaneous} and \cite{miyoshi2011gaussian}
assign a Gaussian prior $\pdf(\beta) = \NormDist(\beta|\beta^\tn{f}, V^\tn{f})$,
where the mean $\beta^\tn{f}$ is a persistence or relaxation of the previous analysis mean,
and where the variance $V^\tn{f}$ is a tuning parameter.
The likelihood is also assumed Gaussian:
$\pdf(\y|\beta) = \NormDist(\hat{\beta}|\beta,\hat{V})$,
where $\hat{\beta} = \hat{\beta}_\I$ in \citet{li2009simultaneous}
and $\hat{\beta} = \hat{\beta}_\R$ in \citet{miyoshi2011gaussian}.
In the former, $\hat{V}$ is a tuning parameter,
while the latter suggests using the variance of $\hat{\beta}_\R$ of \cref{eqn:s_miyoshi11}:
\begin{align}
	\hat{V} =
	\left[ \frac{\trace(\bH \B \bH\tr \R^{-1})/P + 1}
	{\bar{\sigma}^2 \sqrt{P/2}} \right]^2
	\approx
	\left[ \frac{\beta^\tn{f} \bar{\sigma}^2 + 1}
	{\bar{\sigma}^2 \sqrt{P/2}} \right]^2
	\, ,
	\label{eqn:vo_miyoshi11}
\end{align}
where the approximation comes from using $\B \approx \beta^\tn{f} \barB$,
which is consistent with operative assumption that $N=\infty$.
Importantly, \cref{eqn:vo_miyoshi11} has the logical consequence that
the observations are given no weight when
they carry no information, i.e. when
$\bar{\sigma}^2 \rightarrow 0$ or $P \rightarrow 0$.
The method is spatialized by
associating each local analysis domain with its own inflation parameter.
Localization tapering promotes smoothness of the inflation field.

Also working in the framework of a square-root EnKF,
\citet{brankart2010efficient} use diffusive forecasts
for the inflation distribution.
They do not sequentially approximate the posteriors,
instead expressing them explicitly in terms of all of the past innovations,
progressively diffused by a ``forgetting exponent'', $\phi<1$:
\begin{align}
	\pdf(\beta | \y_{1:K})
	&\propto
	\left(\pdf(\beta)\right)^{\phi^K}
	\prod_{k=1}^K
	\NormDist\big(\bbdelta_k \,|\, \bvec{0}, \phi^{k-K} \barC_k(\beta)\big) 
	\, .
	\label{eqn:Brank_k}
\end{align}
The chosen estimate, which maximizes $\pdf(\beta | \y_{1:K})$, is found iteratively.
The requisite evaluations of the cost function is not prohibitive
because the square-root formulation means that diagonalizing matrix decompositions
have already been computed.

\citet{anderson2007adaptive}, working in the framework of the EAKF,
explicitly considers forecasting the (parameters of the) inflation distribution,
but only tests persistence forecasts, i.e. $\pdf(\beta_k | \beta_{k-1}) = \delta(\beta_k - \beta_{k-1})$.
The posteriors (and hence the forecast priors) are approximated by Gaussians:
for a given time,
\begin{align}
	\pdf(\beta|y_i) &\approx \NormDist(\beta|\hat{\beta}_\tn{MAP},V^\tn{a})
	\, ,
	\label{eqn:A07_approx}
\end{align}
which are computed serially for each component $y_i$ of the observation $\y$.
The univariate-observation likelihood
allows the maximum \apost{} (MAP) estimate $\hat{\beta}_\tn{MAP}$ to be found exactly
and without iterations, via a cubic polynomial;
$V^\tn{a}$ is fitted by requiring that \cref{eqn:A07_approx}
be exact for another particular value of $\beta$,
as normalized by its value at $\hat{\beta}_\tn{MAP}$.
\citet{anderson2009spatially}
spatializes the method,
associating each state variable with its own inflation parameter.
Here, correlation coefficients complicate the likelihood,
so that $\hat{\beta}_\tn{MAP}$ must be found via a Taylor expansion of the likelihood
and a resulting quadratic polynomial.

\subsubsection{Prior family}
\citet{anderson2009spatially} also notes that a Gaussian prior
does not constrain the inflation estimates to positive values.
Thus, nonsensical negative estimates will occasionally occur for small innovations.
Imposing some lower cap (bound), e.g., $\hat{\beta}_\tn{MAP}>0$ or even $\hat{\beta}_\tn{MAP}>1$,
is done as a quick-fix.
As \citet{stroud2018bayesian} note, such capping may cause a bias.
However, this bias should be avoided by employing the un-capped values in the averaging.

Instead of ad-hoc mechanisms, using a better tailored prior will pre-empt the problem.
\citet{brankart2010efficient} use an exponential \PDF{}
for the initial prior, but acknowledge that it is not appropriate for small values.

Indeed, a better choice is the inverse-chi-square distribution, $\InvCS$,
also called inverse-Gamma;
it is the conjugate prior to the variance parameter of a Gaussian sample,
and was shown in \cref{sec:Re-deriv} to be intimately linked to inflation.
It has been employed in adaptive filtering by e.g.,
\citet[][]{mehra1970approaches,storvik2002particle},
and in an EnKF contexts by \citet{stroud2007sequential}
and \citet{moha2018inflation},
who used it to enhance the scheme of \citet{anderson2009spatially}.
Similarly, the following section formulates an inflation filter
based on $\hat{\beta}_{\R}$ using $\InvCS$ distributions.

\subsection{Renouncing the Gaussian framework}
\label{sec:gen_purp}

This subsection improves the formulation of the estimator $\hat{\beta}_{\R}$;
in particular, it abstains from assuming Gaussianity for
the distributions for $\beta$.

Recall the definition \labelcref{eqn:bsig_def} of $\bar{\sigma}^2$
and make the approximation $\bH \barB \bH\tr \R^{-1} \approx \bar{\sigma}^2 \I_P$.
This proportionality renders the likelihood symmetric and hence reducible;
indeed, the likelihood \labelcref{eqn:mult_norm_lklhd} can now be written:
\begin{align}
	\pdf(\y|\beta)
	&\appropto
	\NormDist\big(\R^{-1/2} \bbdelta \,\big|\, \bvec{0}, (1 + \bar{\sigma}^2 \beta)\I_P\big)
	\notag
	\\
	&\propto
	(1 + \bar{\sigma}^2 \beta)^{-P/2} e^{-\norm{\bbdelta}_\R^2/2(1 + \bar{\sigma}^2 \beta)}
	\notag
	\\
	&\propto
	\CS\big(\norm{\bbdelta}_\R^2/P \,\big|\, (1 + \bar{\sigma}^2 \beta), P\big)
	\label{eqn:uni_norm_lklhd_3}
	\, .
\end{align}
Remarkably, the value of $\beta$ that maximizes this approximate likelihood is
$\hat{\beta}_{\R}$ of \cref{eqn:s_miyoshi11}.

The likelihood \labelcref{eqn:uni_norm_lklhd_3} may be further approximated
by fitting the following shape to it:
\begin{align}
	\pdf(\y|\beta)
	&\approx
	\CS(\hat{\beta}_{\R}|\beta,\hat{\nu})
	\, .
	\label{eqn:b_lklhd_approx}
\end{align}
Irrespective of the certainty parameter $\hat{\nu}$,
the mode (in $\beta$) is then the same as for \cref{eqn:uni_norm_lklhd_3}, namely $\hat{\beta}_{\R}$.
Also fitting the curvature at the mode to that of \cref{eqn:uni_norm_lklhd_3}
yields $\hat{\nu} = P [\bar{\sigma}^2 \hat{\beta}_{\R}/(1 + \bar{\sigma}^2 \hat{\beta}_{\R})]^2$.
Remarkably,
this yields a variance (cf. \cref{tab:pdfs}) equal to $\hat{V}$ of \cref{eqn:vo_miyoshi11},
except with $\hat{\beta}_{\R}$ in place of $\beta^\tn{f}$.

The benefit of making second approximation \labelcref{eqn:b_lklhd_approx}
in addition to the first \labelcref{eqn:uni_norm_lklhd_3}
is the resulting conjugacy with an $\InvCS$ prior.
Indeed, suppose the forecast prior for the inflation parameter is:
\begin{align}
	\pdf(\beta)
	&=
	\InvCS(\beta | \beta^\tn{f}, \nu^\tn{f})
	\, ,
	\label{eqn:prior_beta}
\end{align}
for some $(\beta^\tn{f}, \nu^\tn{f})$.
The posterior is then:
\begin{align}
	\pdf(\beta| \y) &= \InvCS(\beta|\beta^\tn{a}, \nu^\tn{a})
	\, .
	\label{eqn:b_post_pdf}
\end{align}
where
\begin{subequations}
	\begin{align}
		\label[equation]{eqn:nu_post}%
		\nu^\tn{a} &= \nu^\tn{f} + \hat{\nu} \, ,
		\\
		\label[equation]{eqn:beta_post}%
		\beta^\tn{a} &= (\nu^\tn{f} \beta^\tn{f} + \hat{\nu} \hat{\beta}_{\R})/\nu^\tn{a}
		\, .
	\end{align}
	\label{eqn:b_post}%
\end{subequations}%
This weighted-average update for the parameters is the same as that of the Kalman filter,
except with slightly different meaning to the parameters.
As such,
it constitutes a natural and original derivation of the inflation filter of \citet{miyoshi2011gaussian}.

\subsection{Potential improvements}
\label{sec:pot_improv}
Rather than approximating the likelihood as $\CS$,
an improved approximation could be obtained by
reverting to the likelihood $\pdf(\y|\beta)$ of \cref{eqn:uni_norm_lklhd_3}
and directly fitting $\InvCS(\beta|\beta^\tn{a}, \nu^\tn{a})$
to the posterior $\pdf(\y|\beta) \InvCS(\beta|\beta^\tn{f}, \nu^\tn{f})$
by matching their modes and local curvatures.
As described for \cref{eqn:A07_approx},
fitting posterior criteria is the approach taken by \citet{anderson2007adaptive}.
With the $\InvCS$ prior, though,
an immediate benefit is that of precluding negative and nonsensical
values for $\beta^\tn{a}$.
Moreover, it can be shown that the posterior mode satisfies a cubic equation.
However, the curvature is more complicated; 
alternatives include setting $\nu^\tn{a} = \nu^\tn{f}+\nu$,
where $\nu = \hat{\nu}$
or where $\nu$ is such such that a likelihood
$\CS(\hat{\beta}_{\R}|\beta,\nu)$ would yield the aforementioned cubic-root mode.

The numerical approach offers another set of options:
locating the mode by optimization and computing the curvature by finite differences.
The latter could also be exchanged for the ratio of two points,
as in \citet{anderson2007adaptive}.
If such avenues are pursued,
it is then not necessary to make the approximation of
reducing the likelihood \labelcref{eqn:mult_norm_lklhd} to \labelcref{eqn:uni_norm_lklhd_3};
indeed, it is feasible to find the required statistics with the full likelihood,
provided the pre-computed SVD \labelcref{eqn:svd_nRY}.
However, this is not necessarily advisable, because it yields substantial bias
due to the uncertainty in $\barB$, as discussed in \cref{sec:Single_cycle_bias}.

An alternative approximation is to fit $(\beta^\tn{a}, \nu^\tn{a})$
via the mean and variance of the posterior, as computed by quadrature.
This requires careful implementation to avoid a drift due to truncation errors,
which otherwise accumulate exponentially through the DA cycles.
It is also important to judiciously define the extent of the grid.

Another option is to abandon the parametric approach altogether,
and instead represent the posterior on a grid \citep[e.g.,][]{stroud2018bayesian}.
In the case of a single, global, inflation parameter, as in this paper,
this approach is affordable for any purposeful precision.
Lastly, a Monte-Carlo representation can also be employed \citep[e.g.,][]{frei2012sequential}.

For all variants, a choice must be made as to which point value of $\beta$ to use to inflate the ensemble.
Rather than using the parameter $\beta^\tn{a}$ directly, one can use the mean or the mode (cf. \cref{tab:pdfs}).
Typically, though, $\nu^\tn{a}$ is so large that this does not matter much.
Similarly, although $\hat{\beta}_{\R}$ has a slight bias (\cref{sec:Single_cycle_bias}),
it errs on the side of caution.
For this reason or other,
de-biasing did not generally yield gains in testing by twin experiments.
Therefore, and for simplicity, de-biasing is not further employed.

\subsection{Forecasting}
\label{sec:Forecasting}
The following is a simple and pragmatic
modelling of the forgetting of past information,
achieved by ``relaxing'' towards
some background (and initial) prior, $\InvCS(\beta|\beta_0, 0)$:
\begin{align}
	\pdf(\beta)
	&=
	\InvCS(\beta | \beta^\tn{f}, \nu^\tn{f})
	\, ,
	\label{eqn:beta_f}
\end{align}
where 
\begin{subequations}
	\begin{align}
		\label[equation]{eqn:nu_f}
		\nu^\tn{f} &= e^{-\Delta/L} \nu^\tn{a}
		\, ,
		\\
		\label[equation]{eqn:s_f}
		\beta^\tn{f}   &= e^{-\Delta/L} (\beta^\tn{a}  - \beta_0) + \beta_0 
		\, ,
	\end{align}
	\label{eqn:b_f}%
\end{subequations}
with $\Delta$ as the time between analyses.
The time scale, $L \geq 0$, controls the rate of relaxation,
and could be set as a multiple of the time scale of the model.
Alternatively, it can be set by solving the stationarity condition
$\nu_\infty = e^{-\Delta/L} \nu_\infty + \hat{\nu}$,
derived from \cref{eqn:nu_post,eqn:nu_f}.

The forecast \labelcref{eqn:s_f,eqn:nu_f} was engineered
-- not derived from dynamics.
While the issue is largely academic,
this lack of formality has been perceived as a difficulty
\citep{anderson2007adaptive,anderson2009spatially,sarkka2013non,nakabayashi2017extension}.
The following construes a few possible resolutions.

One possibility is diffusion.
This would yield similar results to \labelcref{eqn:s_f,eqn:nu_f},
though how the $\InvCS$ shape may be maintained is not clear.
However, multiplicative, positive noise seems preferable to avoid negative values.
Computationally, if the parametric distribution is not preserved under the forecast,
then gridded, Monte-Carlo, or kernelized inverse-transform approaches may be employed.
Instead of diffusion, \citet{brankart2010efficient} suggested
exponentiating $\pdf(\beta)$ in the forecast step (cf. \cref{eqn:Brank_k}),
also called ``annealing'' \citep{stordal2015annealed}.
This maintains the $\InvCS$ shape, but is difficult to motivate physically.
Alternatively, the desired effect of ``forgetting''
is perhaps most naturally modelled with an autoregressive model with limited correlation length.
This can be rendered Markovian (to fit with filtering theory) using state augmentation
\citep[][\S 3.4]{durbin2012time}.

It may also be argued that the search for physical dynamics is misguided.
After all, the inflation, $\beta$, is already a hyperparameter.
So why should it be more natural to forecast $\beta$
rather than the hyper-hyperparameters,
$\beta^\tn{a,f}$ and $\nu_k^\tn{a,f}\,$ as in \cref{eqn:s_f,eqn:nu_f}?

\subsection{Specification of the variants shown in the comparative benchmarks}
\label{sec:Specification}

Several of the improvements of \cref{sec:pot_improv} were tested with twin experiments.
The results (not shown) indicate that most schemes, including the original one of \cref{sec:gen_purp},
perform surprisingly well (in terms of filter accuracy)
provided they have reasonable settings of their tunable parameters.
The most plausible explanation is that the inflation filters
are consistent estimators:
as the DA cycles build up, they all eventually converge towards a near-optimal value of inflation.
%
In view of this parity
it seems logical to opt for the simplest scheme,
namely that of \cref{eqn:b_post_pdf,eqn:b_post}.

Similarly, instead of \cref{eqn:nu_f,eqn:s_f},
the numerical experiments use the fixed value $\nu^\tn{f}=10^3$,
corresponding to a variance of approximately $2(\beta^\tn{f})^2/10^3$, according to \cref{tab:pdfs}.
Moreover, instead of fitting the likelihood's certainty, $\hat{\nu}$,
it was simply set to $1$.
Keeping $\nu^\tn{f}$ fixed is suboptimal in the spin-up phase of the twin experiments,
but this part of the experiment is not included in the time-averaged statistics.
Keeping $\nu^\tn{f}$ fixed also foregoes the interesting possibility of
actually having the certainty decrease through an update,
something that will occur with the posterior fitting or non-parametric approaches.
Nevertheless, this simplification was done
(i) to facilitate reproduction;
(ii) because in the twin experiments of this paper,
the models and observational networks are homogeneous,
and therefore using localized and variable $\hat{\nu}$ and $\nu^\tn{f}$ is not crucial;
(iii) it was found that using the fitted $\hat{\nu}$
sometimes yielded worse filter accuracy than keeping it fixed;
(iv) to provide fair comparisons between all of the methods
by equalizing the sophistication of their forecast step.

In addition to the adaptive inflation,
\citet{miyoshi2011gaussian} also uses a fixed inflation of $1.015$.
This was tested and found to make little difference in the experiments herein.
Without directly impacting its distribution,
the value of the inflation actually applied to the ensemble is capped below by $0.9$
for all methods.
However, this clipping never occurred after the spin-up time of the experiment.

The above scheme is practically identical to that of \citet{li2009simultaneous},
the differences having proven largely irrelevant and ``cosmetic'', as discussed above.
It is therefore labelled ``ETKF adaptive''.
Another scheme that was tested, labelled ``EAKF adaptive'',
is that of \cref{eqn:A07_approx}.
Instead of fitting $V^\tn{a}$, however, the value of $V^\tn{f}$ was fixed, and set (tuned) to $0.01$.
As described by \citet{anderson2007adaptive},
the actual inflation applied to the ensemble is damped, using the value $0.9 \hat{\beta}_\tn{MAP}$.
The above two schemes are the established standard in the literature,
and have featured in many operational studies,
although mainly in their spatialized formulations.

The proliferation of tunable ``hyper-hyperparameters'' (the hyperparameters of the inflation filter),
is due to the hierarchical nature of adaptive filters,
and may make the adaptive approach appear counterproductive.
But, considering the breadth of error sources targeted,
it should be recognized that
such methods will necessarily be ad-hoc,
and that the existence of tuning parameters is to be expected.
One should not be dismayed, however,
because the performances of the adaptive filters are largely insensitive to the hyper-hyperparameter settings.
Indeed, intuitively, more abstract parameters (further up the hierarchy) should have less impact,
as is illustrated by the results of \citet{roberts2001infinite}.
Indeed, the given values for the hyper-hyperparameters
(i) were only tuned for a single experimental context,
(ii) seem reasonable,
and (iii) yield satisfactory filter accuracy almost universally across the experiments.
Point (iii) corroborates previous findings \citep[e.g.,][]{anderson2007adaptive,miyoshi2011gaussian},
and suggests that these values may be used \textit{in vivo}.


\section{A hybrid-inflation EnKF-$N$}
\label{sec:hybrid}

\Cref{sec:Re-deriv} showed that
the EnKF-$N$ may yield a form of adaptive inflation, $\alpha_\star$.
However, the EnKF-$N$ is built on \cref{eqn:const_4},
with the assumption that the truth is statistically indistinguishable from the ensemble.
Therefore, it only targets sampling error,
and the prior \labelcref{eqn:lam3} always has the location parameter $1$.
The EnKF-$N$ is therefore not robust in the context of \emph{model} error,
analysed in \cref{sec:lit_review_0}.

If \emph{tuned} inflation is used in concert with the EnKF-$N$,
then the tuned value could be seen as a measure of model error
disentangled from sampling error \citep{bocquet2013joint}.
Here, however, the aim is to hybridize the EnKF-$N$ with an
\emph{adaptive} inflation scheme that estimates an inflation factor, $\beta$, targeted at model error.
Notably, such a scheme has a prior that is time-dependent
and, generally, not with location parameter $1$.
The scheme used is the ``adaptive ETKF'' specified in \cref{sec:Specification}.

Again, the explicit conditioning on the ensemble, $\E$, is here dropped for brevity.
Consider
\begin{align}
	\pdf&(\x,\beta | \y)
	\propto
	\pdf(\y | \x) \pdf(\x | \beta) \pdf(\beta)
	\label{eqn:post_x_beta_12}
	\, .
\end{align}
In contrast with \cref{sec:lit_review_0} and \cref{eqn:prior_x_beta},
$N$ will here not be assumed infinite.
Re-deriving, therefore, the EnKF-$N$ prior \labelcref{eqn:hier_14},
but with \cref{eqn:fawefhawe} in place of \labelcref{eqn:const_4},
reveals that $\pdf(\x | \beta)$ is a scale mixture over $\alpha$,
but now with $\barB$ also scaled by $\beta$,
i.e. $\pdf(\x | \alpha, \beta) = \NormDist(\x|\bx,\alpha\beta\barB)$.
Further, sampling error is assumed independent from model error:
$\pdf(\alpha | \beta) = \pdf(\alpha)$.
Hence,
\begin{align}
	\pdf(\x,\beta | \y)
	\propto
	\pdf(\y | \x) \bigg( \int 
	\pdf(\x | \alpha, \beta) \pdf(\alpha) \diff \alpha \bigg) \pdf(\beta)
	\label{eqn:afwefa}
	\, .
\end{align}
Moving the likelihood inside as in the EnKF-$N$ \labelcref{eqn:prior_marg_y}
yields a mixture over $\pdf(\y, \x | \alpha, \beta)$,
which can be re-factorized to obtain:
\begin{align}
	\pdf(\x,\beta | \y)
	&\propto
	\bigg( \int 
	\pdf(\x | \y, \alpha, \beta) \pdf(\y | \alpha, \beta) \pdf(\alpha) \diff \alpha \bigg) \pdf(\beta)
\end{align}
As in the EnKF-$N$ \labelcref{eqn:post_EB}, the mixture is approximated by empirical Bayes:
\begin{align}
	\pdf(\x,\beta | \y)
	&\appropto
	\underbrace{
		\vphantom{ \int }
	\pdf(\x | \y, \alpha_\star, \beta)
	}_{\approx \pdf(\x | \y, \beta)}
	\bigg( \underbrace{
	\int 
	\pdf(\y | \alpha, \beta) \pdf(\alpha) \diff \alpha}_{\pdf(\y | \beta)}\bigg) \pdf(\beta)
	\label{eqn:post_x_beta_18}
	\, ,
\end{align}
meaning that $\pdf(\x | \alpha_\star, \beta, \y)$ is given by \cref{eqn:post_enkfn},
with the change of variables \labelcref{eqn:CVar_w},
except that the prior covariance is now also scaled by $\beta$,
which also impacts the selection of $\alpha_\star$ through the dual cost function.

The remaining integral of \cref{eqn:post_x_beta_18} is again approximated using a particular value of $\alpha$:
\begin{align}
	\int 
	\pdf(\y | \alpha, \beta) \pdf(\alpha) \diff \alpha
	&\approx
	\pdf(\y | \alpha^\tn{f}, \beta)
	\label{eqn:post_x_beta_22}
	\, .
\end{align}
In this instance, however, the optimizing value of $\alpha$ is not conditioned on $\y$,
and is therefore denoted $\alpha^\tn{f}$.
In practice, $\alpha^\tn{f} = 1$ is used for simplicity.
Thus, $\pdf(\y | \alpha^\tn{f}, \beta)$ becomes the same as in \cref{eqn:b_lklhd_approx},
yielding the posterior of \cref{eqn:b_post_pdf}.

A final approximation is made to decouple the joint posterior:
replacing $\beta$ in the conditional distribution, $\pdf(\x | \y, \alpha_\star, \beta)$,
by $\beta_\star$, some point estimate from $\pdf(\beta | \y, \alpha^\tn{f})$.
This may again be seen as empirical Bayes,
except that $\beta$ is not a latent variable.
Other reasons for only using a single inflation value for the ensemble
are discussed in \cref{sec:hierarchicals}.
The particular point used is the mean:
\begin{align}
	\beta_\star = \frac{\nu^\tn{a}}{\nu^\tn{a}-2} \beta^\tn{a} \, .
	\label{eqn:beta_star}
\end{align}
Thus, \cref{eqn:post_x_beta_18} becomes:
\begin{align}
	\pdf(\x,\beta | \y)
	&\approx
	\pdf(\x | \y, \alpha_\star, \beta_\star)
	\pdf(\y | \alpha^\tn{f}, \beta) \pdf(\beta)
	\notag
	\\
	&\propto
	\pdf(\x | \y, \alpha_\star, \beta_\star)
	\pdf(\beta | \y, \alpha^\tn{f})
	\label{eqn:post_x_beta_24}
	\, .
\end{align}

The algorithm of the analysis update of the EnKF-$N$ hybrid
can be stated as follows.
\begin{enumerate}
	\item Update the general-purpose inflation, $\beta$, according to \cref{eqn:nu_post,eqn:beta_post}.
	\item Conditional on $\beta_\star$ \labelcref{eqn:beta_star}, find the EnKF-$N$ inflation using the dual
		\labelcref{eqn:Dual_def}:
		$\alpha_\star = \argmin_\alpha D(\cN/\beta_\star \alpha)$.
	\item Update the ensemble by the ETKF with a prior inflation of $\alpha_\star \beta_\star$.
		In other words, implement \cref{eqn:Ea_sqrt_zeta} with
		$\cN / \alpha_\star \beta_\star$ in place of $\zeta_\star$.
\end{enumerate}
Note that the second step is the only essential difference to the ETKF adaptive inflation scheme
of \citet{miyoshi2011gaussian}.

The forecast is greatly facilitated by
the decoupling of the posterior \labelcref{eqn:post_x_beta_24},
which means that the ensemble and general-purpose inflation $\beta$ are independent.
As discussed by \cref{sec:Forecasting},
it is pragmatic to separate the forecast of the ensemble
from the forecast of $\beta$,
which should be carried out as in \cref{eqn:b_f}.
Instead, however, for the reasons described in \cref{sec:Specification},
here $\nu^\tn{f} = 10^4$,
and $\beta^\tn{f}$ is set to the previous $\beta^\tn{a}$.
The EnKF-$N$ inflation, $\alpha$, is not forecasted,
as its prior is static.

\section{Benchmark experiments}
\label{sec:Benchmark_experiments}
The standard methods of \cref{sec:Specification}
and the hybrid of \cref{sec:hybrid} are tested with
twin experiments:
a synthetic truth and observation thereof are simulated
and subsequently estimated by the DA methods.
Contrary to the meaning of ``twin'', however,
the setup is deliberately one of model error:
the model provided to the DA system is different from the one
actually generating the truth.

\begin{figure}[htbp]
	\centering
	\includegraphics[width=0.48\textwidth]{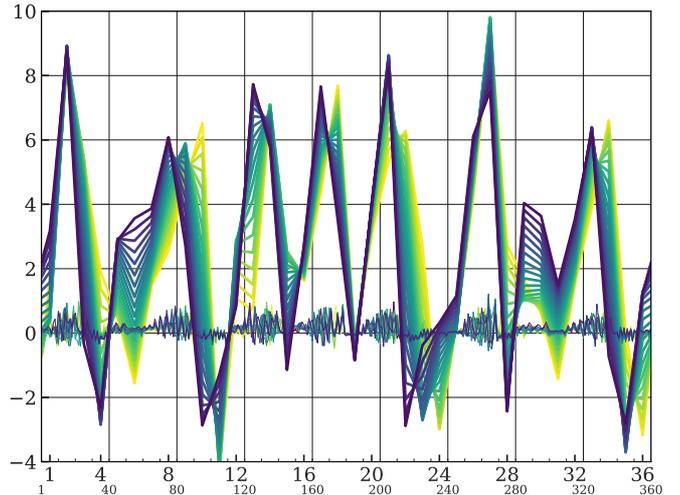}
	\caption[Illustration of two-scale Lorenz model.]
	{
		Illustration of the dynamics of the two-scale Lorenz model
		by 20 consecutive snapshots of the state profile.
		The colour gradation, from light to dark,
		represents the time sequence, with a total span of $0.3$.
		The large-scale $\x$ field with a mean value of 2.4 can be seen moving left.
		The small-scale $\z$ field with a mean value of 0.1 is moving right;
		its variations are faster, but of less amplitude.
		The abscissa denotes the indices of (upper) $\x$ and (lower) $\z$.
	}
	\label{fig:LorenzUV}
\end{figure}

The main system used is the
two-scale/layer Lorenz model \citep{lorenz1996predictability,lorenz2005designing}.
It constitutes a surrogate system for synoptic weather,
exhibiting similar characteristics,
and enables studying the impact of unresolved scales on filter accuracy.
The autonomous dynamics are given by:
$
	\psi_i^\pm(\bu)
	=
	u_{i \mp 1} (u_{i \pm 1} - u_{i \mp 2}) - u_{i}
$,
where the indices apply periodically. 
Then,
\begin{align}
	\dd{x_i}{t}
	&=
	\psi_i^+(\x)
	+ F - h \frac{c}{b}  \sum_{j=1}^{10} z_{j + 10(i-1)}
	\, ,
	\label{eqn:Lor2X}
	\\
	\dd{z_j}{t}
	&=
	\frac{c}{b} \psi_j^-(b\z)
	+ 0
	+ h \frac{c}{b} x_{1+(j-1)/\!/10}
	\, ,
	\label{eqn:Lor2Y}
\end{align}
for $i = 1,\ldots,36$, $j = 1,\ldots,360$, and
where $/\!/$ means integer division.
Unless otherwise stated, 
the constants are set as in \citet{lorenz1996predictability}:
time-scale ratio: $c=10$,
space-scale ratio: $b=10$,
coupling: $h=1$,
forcing: $F=10$.
The resulting dynamics,
illustrated in \cref{fig:LorenzUV},
are chaotic and have a leading Lyapunov exponent of 1.3775 \citep{mitchell2014accounting}.

The truth is composed of both fields:
$\begin{bmatrix} \x, \z \end{bmatrix}$.
By contrast, 
the model provided to the DA methods is a truncated version of the full one:
\begin{align}
	\dd{x_i}{t}
	&=
	\psi_i^+(\x)
	+ F - [A + B x_i]
	\, , && i = 1,\ldots,36.
\end{align}
The term in the brackets parameterize (compensate for)
the missing coupling to the small scale.
Ahead of each experiment, the constants $A$ and $B$ are determined
by linear regression to unresolved tendencies,
as described by \citet{wilks2005effects}.
The error parameterization removes the linear bias of the truncated model;
this has been done because inflation is not well suited to deal with systematic errors.
Higher-order parameterizations were tested and found to yield little improvement to the filter accuracies,
while a $0$-order parameterization yielded too much model error,
overly dominating the dynamical growth in the filter error.

Another source of model error is that
the full model is integrated with a time step of $0.005$, as necessitated by stiffness \citep{berry2014linear},
while the truncated model uses $0.05$ to lower computational costs.
However, this source of error was found to be negligible compared to the truncation itself.

The time between observations is $\Delta = 0.15$.
Direct observations are taken of the full $\x$ field with error covariance $\R = \I_M$.
There is no model noise: $\Q = \mat{0}$.
The filters are assessed by their accuracy
as measured by root-mean squared error:
\begin{align}
	\RMSE = \sqrt{\frac{1}{M} \norm{\x - \bx}^2}
	\, ,
	\label{eqn:RMSE_3}
\end{align}
which is recorded immediately following each analysis.
This instantaneous RMSE
is then averaged in time (3300 cycles following a spin-up of 40 cycles),
and over 32 repetitions of each experiment.
A table of RMSE averages is compiled for a range of experiment settings
and plotted as curves for each method.
The figures also present (thin, dashed lines) the root-mean variances (RMS spread) of each method.
All of the experiment benchmark results can be reproduced using Python-code scripts hosted online at
\url{https://github.com/nansencenter/DAPPER/tree/paper_AdInf}.

\begin{figure*}
	\subfloat[
		From the two-scale system, using $N=20$.
	]
	{\label{fig:LUV_F}\includegraphics[width=0.495\textwidth]{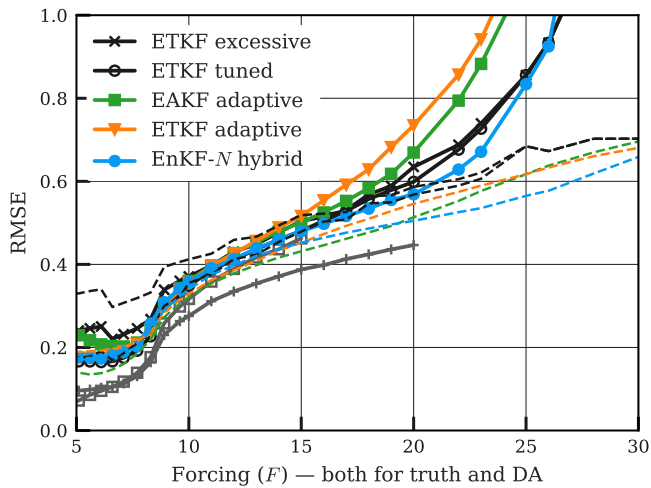}}
	\hspace{\stretch{1}}
	\subfloat[
		From the two-scale system, using $N=20$.
	]
	{\label{fig:LUV_c}\includegraphics[width=0.495\textwidth]{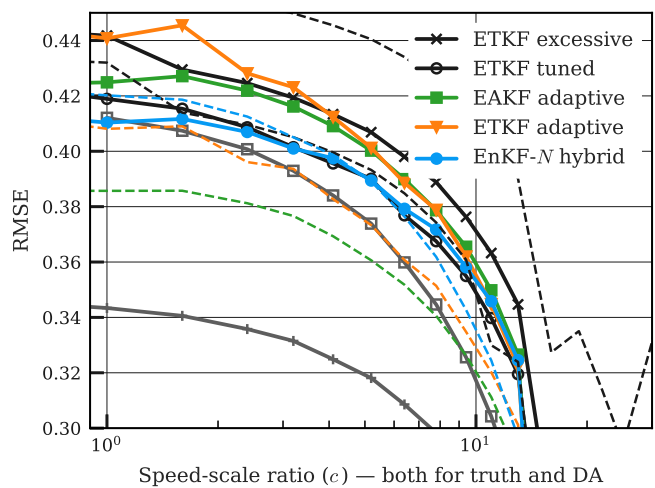}}
	\\
	\subfloat[
		From the single-scale system, using $N=20$.
	]
	{\label{fig:L95_FTr}\includegraphics[width=0.495\textwidth]{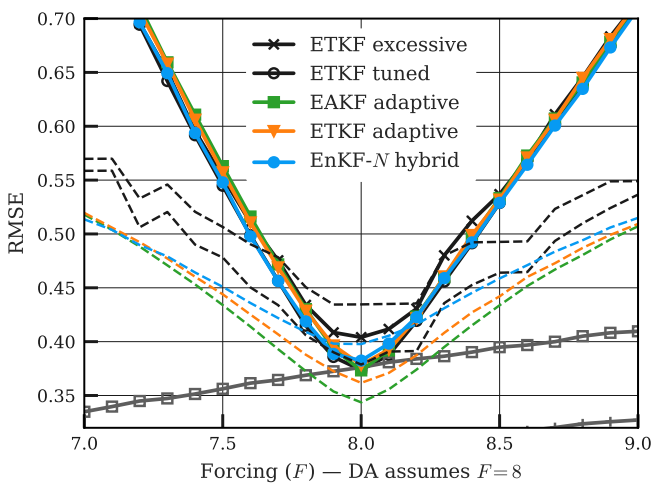}}
	\hspace{\stretch{1}}
	\subfloat[
		From the Lorenz-63 system, using $N=3$.
	]
	{\label{fig:L63_Q}\includegraphics[width=0.495\textwidth]{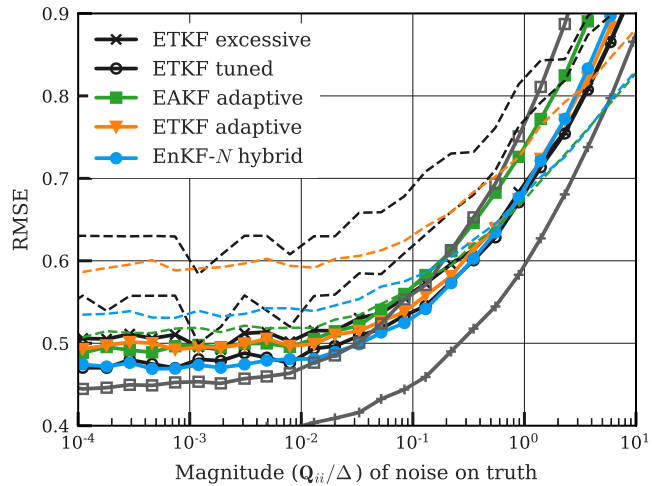}}
	\caption[
		Filter accuracy benchmarks as a function of various control variables.
	]
	{
		Accuracy benchmarks of the adaptive inflation filters,
		plotted as functions of various control variables.
		Also included in the plots is the RMS spread, plotted as thin, dashed lines.
		For perspective, two baselines are provided, in grey.
		These are obtained with the full (i.e. perfect) model using the pure EnKF-$N$:
		one with the same ensemble size as the adaptive filters (marker: $\square$),
		and one with $N=80$ (marker: $+$).
		Among the adaptive methods,
		the proposed hybrid EnKF-$N$ (blue) scores the lowest RMSE averages
		nearly systematically across all contexts, albeit by a moderate margin.
	}
	\label{fig:benchmarks}
\end{figure*}

%

%

\Cref{fig:LUV_F} shows benchmarks obtained with the two-scale system as a function of the forcing, $F$.
The increasing RMSE averages of all filters
reflect the fact (not shown) that the system variability and chaoticity both increase with $F$.
The same applies for decreasing $c$ in \cref{fig:LUV_c}, where $F$ is fixed at $10$.
Note that all of the adaptive filters are largely coincident at $F=10$ and $c=10$,
with RMSE scores almost as low as fixed, tuned inflation.
This is because the hyper-hyperparameters for each method,
described in \cref{sec:Specification},
were tuned at this point (and this point only).
By contrast, the fixed inflation of the ``ETKF tuned'' filter
is determined for each experiment setting by selecting for the lowest RMSE
among 40 inflation values between $0.98$ and $3$, most of which are close to $1$.

It is not surprising that tuning an adaptive filter
will make it about as accurate any other.
The objective, however, is to avoid tuning.
In that regard, it is surprising is how well all of the adaptive filters perform overall.
Indeed, except for the fairly extreme contexts of $F>15$ or $c<4$,
the difference in RMSE is small
in the sense that the adaptive filters are all superior to ``ETKF excessive'':
a fixed-inflation filter with a suboptimal inflation factor that adds $0.1$ to the optimal value.

Benchmarks were also obtained with the single-scale system,
where both the truth and the DA systems are given by:
\begin{align}
	\dd{x_i}{t}
	&=
	\psi_i^+(\x)
	+ F
	\, , && i = 1,\ldots,40,
\end{align}
and there is no $\z$ field.
As in \citet{anderson2007adaptive},
the model error consists in using
a different value of $F$ for the truth than for the DA system.
The setup is otherwise repeated from above.
As shown in \cref{fig:L95_FTr},
the benchmarks plotted as a function of $F$ are V-shaped,
with the lowest scores obtained in the absence of error ($F=8$).
The adaptive filters score very similar RMSE averages,
which are generally significantly in excess of the RMS spread scores.
The mismatch can be explained by the well known bias-variance decomposition
of RMSE, and the fact that the model error contains significant bias.
The presence of bias is also a likely cause for the closeness of the
RMSE averages of the adaptive methods,
because inflation is not well suited to treat bias.

Tests were also run with the 3-variable Lorenz-63 system
\citep{lorenz1963deterministic},
where the model error consists in adding independent white noise to the truth.
The setup is the same as above,
except that $\R = 2 \I_3$.
\Cref{fig:L63_Q} shows the corresponding benchmarks.
These are obtained with a small ensemble ($N=3$);
using a larger ensemble, the relative advantage of the hybrid disappears.

\begin{figure}[htbp]
	\centering
	\includegraphics[width=0.48\textwidth]{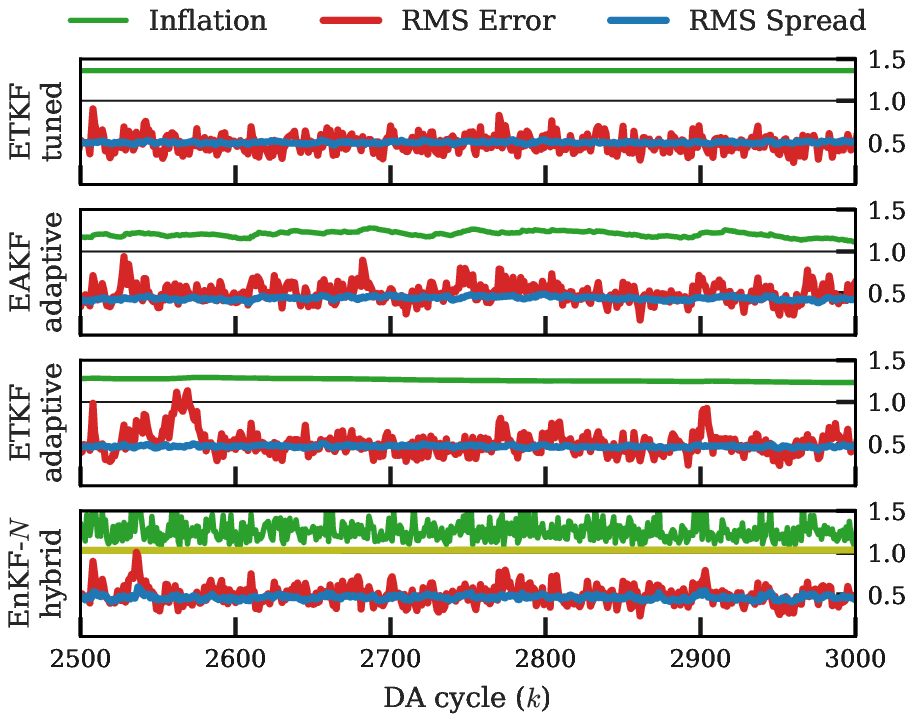}
	\caption[]
	{
		Illustration of the statistics from the twin experiments
		by a segment of the typical time series.
		Generated with the two-scale Lorenz model, with $F=16$.
		In the panel of the hybrid EnKF-$N$,
		the second inflation line (yellow) indicates the value of $\beta^\tn{a}$,
		and averages $1.02$.
	}
	\label{fig:stats_series}
\end{figure}

The hybrid EnKF-$N$ obtains slightly superior accuracy relative to
the adaptive ETKF and EAKF for nearly all experiments.
This is as expected from theory: separate, dedicated treatment of
sampling and model errors yield improved accuracy.
The practical advantage of the hybrid is illustrated in the
time series of \cref{fig:stats_series}.
Notably, the inflation of the hybrid EnKF-$N$ has much more volatility
(shorter time scale).
This is made possible by the static prior which ``anchors'' the inflation to $1$.
By contrast, similar volatility in the adaptive ETKF and EAKF would
require much more lenient settings of $\nu^\tn{f}$ (or $V^\tn{f}$),
which would yield excessive longer-term volatility (i.e. variance).

On the other hand, the volatility also means that larger spikes in the inflation will occur.
Since inflation is not a physical or especially gentle way of increasing spread,
this could potentially cause trouble.
For example, the pure EnKF-$N$ with the full model and large $F$ values,
blows up due to stiffness, as illustrated by the grey curves in \cref{fig:LUV_F}.
This could have been prevented by increasing (doubling) its certainty parameter,
something that would also slightly improve its accuracy across all of the experiment settings.

Different hyper-hyperparameter values for the adaptive EAKF and ETKF
will penalize their RMSE scores for some settings,
and reward it for other settings.
This sensitivity was observed to be much reduced for the hybrid,
which is in line with the principal objective:
to avoid tuning across a multitude of contexts.

A secondary objective is to obtain improved accuracy compared to fixed, tuned inflation,
as has been previously observed for the pure EnKF-$N$
in the perfect-model context \citepalias{bocquet2012combining}.
\Cref{fig:benchmarks} shows that this is sometimes achieved,
but by a very small margin.

Further experiments (not shown) were carried out,
using different ensemble sizes and other types of model errors.
The trends were similar to the benchmarks already shown,
but typically with less relative difference between the filters.

\section{Concluding remarks}
\label{sec:Discussion}

This paper has developed an adaptive inflation scheme
as a hybrid of ($\alpha$) the finite-size ensemble Kalman filter inflation
\citepalias[the EnKF-$N$ of][]{bocquet2011ensemble,bocquet2012combining,bocquet2015expanding}
and ($\beta$) the inflation estimation conventionally associated with the ensemble transform Kalman filter (ETKF).
In so doing, it has provided several novel theoretical insights on the EnKF and adaptive inflation estimation.

The first part of the paper is focused on idealistic contexts,
with sampling error being the main concern.
Using two univariate toy experiments,
\cref{sec:Two_scalar_experiments} illustrated the generation of sampling error
by nonlinearity, as predicted by \cref{sec:nonlin_sampling_err}.
\Cref{sec:catalogue} then discussed the circumstances for inflation,
cataloguing them according to linearity, stochasticity, and ensemble size.
The discussion revealed
why sampling errors are attenuated in the linear context,
why the choice of normalization factor (e.g., $\frac{1}{N-1}$) is not crucial,
and also touched on topics such as ensemble collapse and filter divergence.
Next, \cref{sec:Sketch} gave a birds-eye view of the EnKF-$N$,
showing how (e.g., empirical Bayes) and why (e.g., feedback) it works.
The following sections filled in the details;
in particular,
\cref{sec:Prior_as_scale} showed how the effective prior
reduces to a Gaussian scale mixture,
again demonstrating the relationship between sampling error and inflation.
The mixture parameter, $\alpha$, is shown to be $\InvCS$.
\Cref{sec:saddle} derived a saddlepoint form
to retain the inflation-explicit expressions all the way up to the posterior.
Without recourse to Lagrangian duality theory,
\cref{sec:posterior} then finalized the re-derivation of the (dual) EnKF-$N$
by showing how the mode of the posterior may be found by optimizing for the inflation factor, $\alpha$.

In contrast to the above, \cref{sec:lit_review_0} is focused on model error,
neglecting sampling error.
A formal and unifying survey of the existing adaptive inflation literature is presented;
particular attention is given to the schemes
conventionally associated with the EAKF and the ETKF ($\hat{\beta}_{\R}$).
The ETKF scheme is given a new and natural derivation in \cref{sec:gen_purp},
again yielding the $\InvCS$ distribution.
Several potential improvements, some novel, were discussed in \cref{sec:pot_improv},
but generally found to be less rewarding than hoped for.
\Cref{sec:more_on_marg} gives some new results on biases, including the maximum likelihood estimator.
The survey is supplemented by
\cref{sec:Joint_xs_estim} on joint schemes,
including a suggestion for inflation estimation by variational Bayes.
\Cref{sec:Forecasting} commented on the forecasting of hyperparameters such as the inflation parameter, $\beta$.

Combining the above, \cref{sec:hybrid} developed
a hybrid between the EnKF-$N$ and the adaptive ETKF inflation scheme.
The hybrid employs two inflation factors, $\alpha$ and $\beta$,
separately targeting sampling and model error, respectively.
The EnKF-$N$ component ($\alpha$) adds
negligible computational cost and no further tuning parameters
to the ETKF and its adaptive inflation ($\beta$),
yet increases the inflation volatility and thus ability.
The experiments of \cref{sec:Benchmark_experiments} showed
that the hybrid generally yields similar filter accuracy as fixed inflation,
even in bias-dominated contexts,
but without the costly need for tuning.
It also yields improved filter accuracy in comparison with the standard, pre-existing
adaptive inflation schemes of the ETKF and the EAKF.

Unless the ensemble size was small and the context strongly nonlinear,
however, the gains were found to be relatively modest,
as was the difference in between the existing methods.
This is somewhat surprising in view of
the essential importance of inflation in many configurations of the EnKF.
Part of the explanation may be that, as a hyperparameter,
the accuracy of the inflation estimates is not as important as that of the (primary) state variables and that,
instead, the main importance of the inflation scheme consists in its capacity to avoid divergence occurrences,
which is a matter of a more boolean character.
Another cause is that the inflation estimates converge and become nearly constant within a relatively short span of time,
and that these asymptotic estimates are sufficiently accurate for all of the methods.

While the experimental results clearly demonstrated
the improvements of the hybrid adaptive inflation scheme,
extrapolating these findings to other, larger applications is non-trivial.
Spatialization of the inflation parameter will likely be necessary;
it may be implemented without considerable complexity as in
\citet{miyoshi2011gaussian}.
Still, the relative modesty of the above experimental results
does not promise great, general gains.
On the other hand it suggests the conclusion,
aided by the rigour and scope of this study,
that further sophistication of single-factor adaptive inflation estimation schemes
is unlikely to yield significant, further improvements.

\appendix

\section{Standard distributions}
\label{sec:standard_pdfs}
\Cref{tab:pdfs} specifies the distributions in use in this paper.
The following properties are useful.

\newcommand{\myNewLine}[0]{\\[0.4em]}
\begin{table*}[htbp]
	\caption{
		Parametric probability distributions.
		As elsewhere in the paper,
		$\bb,\x \in \Reals^M$,
		$\B,\bS \in \CovSpace$,
		$s,\beta>0$, and it is assumed that $\nu>M$.
		The constants are
		$\CNorm = (2\pi)^{-M/2}$,
		$\Ctdist = \frac{\Gamma(\frac{\nu+M}{2})}{(\pi\nu)^{M/2} \Gamma(\nu/2)}$,
		$\CWish = \frac{\nu^{\nu/2}}{2^{{\nu M}/{2}} \Gamma_M(\nu/2)}$,
		and $\CCS=\CWish$ with $M=1$.
		The (not listed) variance of element $(i,j)$ of $\B$
		with the Wishart distribution is $(s_{ij}^2 + s_{ii}s_{jj})/\nu$,
		where $s_{ij}$ is element $(i,j)$ of $\bS$.
		The variances of the inverse-Wishart distribution are asymptotically, for $\nu \to \infty$, the same.
	}
	\label{tab:pdfs}
	\begin{center}
	\begin{tabular}{llllll}
		\toprule
		\textbf{Name} & \textbf{Symbol} & \textbf{Probability density function}
		& \textbf{Mean} & \textbf{Mode} & \textbf{(Co)Var} \\
		\midrule
		Gauss./Normal &
		$\NormDist(\x | \bb, \B)$ &
		\makebox[2em][l]{$=\CNorm$}
		$\abs{\B}^{-1/2} \exp \big(\mahalf \norm{\x-\bb}_{\B}^2)$ &
		$\bb$ & $\bb$ & $\B$
		\myNewLine
		$t$ distribution &
		$\tdist(\x|\nu;\bb,\B)$ &
		\makebox[2em][l]{$=\Ctdist$}
		$\abs{\B}^{-1/2}\big(1+\frac{1}{\nu}\norm{\x-\bb}_{\B}^2\big)^{-(\nu+M)/2}$ &
		$\bb$ & $\bb$ & $\frac{\nu}{\nu-2} \B$
		\myNewLine
		Wishart &
		$\Wishart(\B | \bS,  \nu)$ &
		\makebox[2em][l]{$=\CWish$}
		$\abs{\bS}^{-\nu/2} \abs{\B}^{{(\nu - M - 1)}/{2}} e^{-\trace(\nu \B \bS^{-1})/2}$ &
		$\bS$ & $\frac{\nu - M -1}{\nu} \bS$ &
		\myNewLine
		Inv-Wishart &
		$\InvWish(\B | \bS,  \nu)$ &
		\makebox[2em][l]{$=\CWish$}
		$\abs{\bS}^{\nu/2} \abs{\B}^{-{(\nu + M + 1)}/{2}} e^{-\trace(\nu \bS \B^{-1})/2}$ &
		$\frac{\nu}{\nu-M-1} \bS$ & $\frac{\nu}{\nu+M+1} \bS$ &
		\myNewLine
		Chi-square &
		$\CS(\beta | s, \nu)$ &
		\makebox[2em][l]{$=\CCS$}
		$s^{-\nu/2} \beta^{\nu/2 - 1} e^{-{\nu \beta}/{2 s}}$ &
		$s$ & $\frac{\nu-2}{\nu}s$ & $2 s^2/\nu$
		\myNewLine
		Inv-chi-square &
		$\InvCS(\beta | s, \nu)$ &
		\makebox[2em][l]{$=\CCS$}
		$s^{\nu/2}\beta^{-\nu/2-1} e^{- {\nu s}/{2 \beta}}$ &
		$\frac{\nu}{\nu-2}s$ & $\frac{\nu}{\nu+2}s$ & $\frac{2 (\nu s)^2}{(\nu-2)^2(\nu-4)}$ \\
		\bottomrule
	\end{tabular}
	\end{center}
\end{table*}
%
\begin{property}
	\label{prop:gamma}
	The (``scaled'') chi-square distributions are equivalent to the Gamma distributions:
	\begin{align}
		\CSN^{\pm 2}(\beta | s, \nu) &= \tn{Gamma}^{\pm 1}(\beta|\nu/2,\nu s^{\mp 1} /2) \, ,
	\end{align}
	where the switch sign $\pm$ has been used to represent both the regular and inverse distributions.
The $\CSN$ parameterization has been preferred
for the parameter interpretations offered by \cref{prop:IG_CV_G},
and the notational simplicity of \cref{prop:scalar,prop:CS}.
\end{property}
\begin{property}
	\label{prop:IG_CV_G}
	Asymptotic normality.
	If $\beta \sim \CSN^{\pm 2}( s, \nu)$,
	then the distribution of $\sqrt{\nu} (\beta - s)$ converges to $\NormDist(0, 2 s^2)$
	as $\nu \rightarrow \infty$.
\end{property}

	Since it describes the sum of squared Gaussians,
	the asymptotic result for $\CS$ is a consequence of the central limit theorem.
	The result for $\InvCS$
	can be shown by through the pointwise convergence of the \PDF{} of $\sqrt{\nu} (\beta - s)$,
	normalized by its value at 0.

	Note that the same limit would have applied if $\beta \sim \NormDist(s, 2 s^2/\nu)$.
	This shows that $s$ plays the role of a location parameter in $\CSN^{\pm 2}( s, \nu)$,
	while $2 s^2/\nu$ plays the role of variance,
	and explains why ``certainty'' is preferred to ``degree of freedom'' for $\nu$ in this paper.

\begin{property}
	\label{prop:scalar}
	In the univariate case ($M=1$),
	\begin{align}
		\WishartN^{\pm 1}(\beta | s,  \nu) &= \CSN^{\pm 2}(\beta | s, \nu) \, .
	\end{align}
\end{property}
\begin{property}
	\label{prop:CS}
	Reciprocity. With $t = 1/\beta$,
	\begin{align}
		\pdf(\beta)  &= \InvCS(\beta|s, \nu) \notag \\
		\text{iff. } \pdf(t) &= \CS(t|1/s, \nu) \, .
	\end{align}
\end{property}
\begin{property}
	\label{prop:Wish}
	Reciprocity. With $\T = \B^{-1}$,
	\begin{align}
		\pdf(\B) &= \InvWish(\B|\bS, \nu) \notag \\
		\text{iff. } \pdf(\T) &= \Wishart(\T|\bS^{-1}, \nu) \, ,
	\end{align}
	as follows by the change of variables and the Jacobian
	$\abs{\T}^{-(M+1)}$ \citep[][\S 2.1]{muirhead1982aspects}.
\end{property}
\begin{property}
	\label{prop:WishChi2}
	Let $\bu \neq \bvec{0}$ be any $M$-dimensional vector,
	or an (almost never zero) random vector.
	If $\T \sim \Wishart(\bS,\nu)$ is independent of $\bu$,
	then
	\begin{align}
		\frac{\bu \tr \T \bu}{\bu \tr \bS \bu} &\sim \CS(1,\nu)
		\, .
	\end{align}
	Moreover, this statistic is also independent of $\bu$.
	Proof: theorem 3.2.8 of \citet[][]{muirhead1982aspects}.
\end{property}

\section{Nonlinearity and sampling error}
\label{sec:nonlin_sampling_err}
This discussion complements that of \cref{sec:Two_scalar_experiments}.

\subsection{Why does nonlinearity generate sampling error?}
\label{sec:Why_NSE}
First, consider what is meant by ``sampling error''.
A sample does not \emph{per se} have a sampling error;
it is by definition random, i.e. subject to variation.
By contrast, estimators, or rather their realized estimates, have sampling error:
the difference between the estimate and its expected value.
By extension, any statistic (any function of the sample) may be said to have sampling error;
for simplicity, however, the discussion below is limited to the non-central sample moments,
i.e. $\hat{\mu}_m(\{x_n\}_{n=1}^N) = N^{-1}\sum_{n=1}^N x_n^m$, in the univariate case.
If the sample is drawn from the same distribution as $x$,
then $\hat{\mu}_m$ is an unbiased estimate of
the $m$-th moment of $x$, i.e. $\mu_m = \Expect[x^m]$,
and the sampling error is the difference:
\begin{align}
	\tn{Error}_m &= \hat{\mu}_m - \mu_m
	\, .
	\label{eqn:samp_err_m}
\end{align}

It is a well known property of the Kalman filter that
the covariance does not depend on the mean.
In the forecast step, this is due to the fact that
their evolutions are entirely \emph{decoupled}.
Indeed, in the case of linear dynamics ($d=1$),
the $m$-th forecast moment is given by:
$\mu_m^\tn{f} = \M^m \mu_m $,
where $\M$ is the (scalar) linear model:
$x^f = \M x$.
For nonlinear forecast dynamics $\dynm$, however,
the moments will be coupled through $\dynm$.
For example, if (locally to the support of $\pdf(x)$)
the model $\dynm$ can be represented by a polynomial of degree $d$,
then the $m$-th moment of the random variable $\dynm(x)$ is a linear combination
of moments of $x$ of order $1$ through $m d$:
\begin{align}
	\mu_m^\tn{f}
	&=
	\sum_{i=1}^{m d}
	C_{m,i}
	\mu_i
	\, .
\end{align}
Thus, for $d>1$ the moments get mixed and, in particular,
impacted by moments of higher order.
This is known as the ``closure problem'' \citep[e.g.,][\S 29]{lewis2006dynamic}.

A similar analysis reveals that the same coupling takes place for the sample moments, $\hat{\mu}_m$.
Therefore the sampling errors are also coupled:
\begin{align}
	\tn{Error}_m^\tn{f}
	&=
	\sum_{i=1}^{m d}
	C_{m,i}
	\tn{Error}_i
	\, .
	\label{eqn:error_mix}
\end{align}
But an $N$-sized ensemble can only match $N$ moments,
e.g., $\tn{Error}_i = 0$ for $i=1,\ldots,N$,
and so there will always be some sampling error present for $i>N$.
Then, by the mixing of \cref{eqn:error_mix},
this error will cascade into the lower-order forecast errors.
In summary, nonlinearity causes sampling error
in (e.g.) the mean and covariance
by pulling in the inevitable (with finite $N$)
sampling error from higher-order moments.

The above analysis is concerned with the \emph{generation} of sampling error.
Another reason for sampling error in the context of nonlinearity
is that chaos prevents its \emph{elimination}
by limiting the effect of far-past observations
(as opposed to the linear case illustrated in \cref{fig:NonLin_yet_Gaussian},
where the initial sampling error is quickly attenuated).
It is not immediately clear whether this is a separate cause
or, rather, a different perspective on the same phenomenon.

\subsection{A nonlinear model preserving $\NormDist$}
\label{sec:MNonlin}
The nonlinear model of \cref{sec:Two_scalar_experiments} was
designed using ``inverse transform sampling''.
It is specified by:
$\dynNonL(x) = \sqrt{2} \CdfNorm^{-1} \big( \CdfCS (x^2) \big)$,
where $\CdfCS$ is the cumulative distribution function (CDF) for $\CS(1,1)$,
and $\CdfNorm^{-1}$ is the inverse CDF for $\NormDist(0,1)$.
	For context, note that $\dynNonL$:
	(i) is V-shaped, with a singularity at $0$,
	(ii) is closely approximated as
	$\dynNonL(x) \approx \sqrt{2}\{0.88\abs{x} + 0.23 \log(x^2) -0.4 \}$,
	(iii) applied to a density symmetric about 0, it may be visualized
	as folding it up in the middle before smearing it back out again,
	(iv) may be generalized to higher dimensionality
	by expressing $\x$ in polar coordinates,
	since $\pdf(\norm{\x}^2) = \CS(\norm{\x}^2|M,M)$
	if $\pdf(\x) = \NormDist(\x|\bvec{0},\I_M)$.

Presumably, any nonlinear model that preserves Gaussianity must include a singularity.
This renders the example using $\dynNonL$ to generate sampling error without non-Gaussianity
somewhat artificial,
but does not jeopardize the utility of considering the two issues separately.

\section{Joint state-covariance estimation}
\label{sec:Joint_xs_estim}

As mentioned immediately above \cref{sec:Marginal_s_estim},
the joint approach approximates $\pdf(\x,\beta|\y)$ simultaneously in $\x$ and $\beta$.
Thus, their analyses impact each other.
This is particularly relevant for state-inflation (or state-covariance) estimation problem,
because of the non-Gaussianity of $\pdf(\x,\beta|\y)$.

\subsection{Variational methods}
\label{sec:variationals}
A common approximate solution to this non-Gaussianity
is to use variational methods for parametric fitting.
As detailed below, this leads to iterative schemes where
the update to the hyperparameter, $\beta$,
is computed in terms of the updated ensemble (for $\x$),
and \vicev{}.
Some of the literature below is concerned with estimating $\R$ (jointly with $\x$),
but is still pertinent by the proximity of the problem to that of estimating $\B$.

\citet{sarkka2009adaptiveVB}
introduce the Variational Bayes (VB) method in the framework of the Kalman filter.
\citet{nakabayashi2017extension} extend it to the EnKF.
The VB method imposes an approximate posterior with
two factor distributions, $\qdf(\x|\y) \qdf(\R|\y)$,
fitted by minimizing the
Kullback-Leibler divergence from the correct posterior, $\pdf(\x,\R|\y)$.
This yields the condition that each factor be
the (geometric) marginal of $\pdf(\x,\R|\y)$ with respect to the other,
i.e. two coupled \PDF{} equations.
Independence is assumed for the prior:
$\pdf(\x,\R) = \NormDist(\x|\bx,\barB)\InvWish(\R|\R^\tn{f}, \nu^\tn{f})$.
It is then shown that the distributions of the approximate posterior are again 
$\NormDist$ and $\InvWish$,
with parameters computable by fixed point iteration.
Iteration $i+1$ consists of the EnKF equations
using the $i$-th estimate of $\R^\tn{a}$ to compute
the analysis mean, $\bx^\tn{a}$, and covariance, $\barPa$,
whose updated values are then used to in the update
$\R^\tn{a} = \big\{\nu^\tn{f} \R^\tn{f} + \hat{\R} \big\}/\nu^\tn{a}$
where $\nu^\tn{a} = \nu^\tn{f} + 1$ and
\begin{align}
	\hat{\R}
	&=
	(\y - \bH \bx^\tn{a})(\y - \bH \bx^\tn{a})\tr + \bH \barPa \bH\tr
	\, .
	\label{eqn:VB_Ra}
\end{align}

\citet{ueno2016bayesian}, also estimating $\R$ in the framework of the EnKF,
use the expectation maximization (EM) method to
maximize the marginal posterior $\pdf(\R|\y)$, with an $\InvWish$ prior.
The expectation is over $\x$, given $\y$ and the current estimate of $\R$.
The empirical distribution is assumed for the prior:
$\pdf(\x) \approx N^{-1}\sum_n \delta(\x-\x_n)$,
yielding iterations involving weighted statistics.
Our investigation indicates that if, instead,
the standard EnKF assumption had been used:
$\pdf(\x) \approx \NormDist(\x|\bx,\barB)$,
then the method would have yielded iterations as in \cref{eqn:VB_Ra}.
This is unsurprising in view of the close connection between EM and VB.

An original result is obtained by applying the VB method to estimate the inflation parameter.
Consider the prior
$\pdf(\x,\beta) = \NormDist(\x|\bx,\beta\barB) \InvCS(\beta|\beta^\tn{f},\nu^\tn{f})$;
note that $\x$ and $\beta$ are not assumed independent.
It can then be shown that the resulting VB scheme consists of using the $i$-th iterate of $\beta^\tn{a}$
to compute the $i+1$ iterates of $\bx^\tn{a}$ and $\barPa$
which, in turn, are used to compute the $i+1$ iterate
$\beta^\tn{a} = \big\{\nu^\tn{f} \beta^\tn{f} + \hat{\beta}\big\}/\nu^\tn{a}$
where $\nu^\tn{a} = \nu^\tn{f} + M$ and
\begin{align}
	\hat{\beta}
	&=
	\norm{\bx^\tn{a}-\bx^\tn{f}}^2_{\barB} + \trace(\barPa \barB^{-1})
	\, ,
	\label{eqn:VB_sa}
\end{align}
which should be computed in the ensemble subspace if $N \leq M$.
\Cref{eqn:VB_sa} may be interpreted using
the trigonometric relations of \citet{desroziers2005diagnosis}.
Numerical experiments indicate that the bias of this VB method (with a flat prior)
is significantly higher than for $\hat{\beta}_\R$, typically also with a larger variance.
It seems likely that the reason is similar to that of $\hat{\beta}_\tn{ML}$,
analysed in \cref{sec:Single_cycle_bias}.


\subsection{Hierarchical methods}
\label{sec:hierarchicals}
Instead of the variational approach,
a more principled approach is to represent
the marginal $\pdf(\beta|\y)$ with a Monte-Carlo sample.
This yields a hyper-ensemble
of distributions $\pdf(\x|\beta,\y)$ and their ensembles.
The label ``hierarchical'' is sometimes reserved for adaptive filters
treating the hyperparameter in this more Bayesian (i.e. full-\PDF{}) manner.
However, the number of realizations quickly becomes exorbitant.
On the other hand, if only a single sample is drawn from $\pdf(\x|\beta,\y)$ for each $\beta$
(i.e. not using an ensemble of ensembles),
then it seems especially prone to spurious correlations.
Another criticism is that it might cause discrete jumps between localization domains.

\citet{myrseth2010hierarchical} provided one example of a hierarchical EnKF,
but without conditioning the hyperparameter on $\y$,
without accounting for model error,
and with a forecast of (the hyperprior of) $\B$ that decays towards $\mat{0}$.
\citet{tsyrulnikov2015hierarchical} present an ambitious hierarchical EnKF,
explicitly considering both model and sampling error.
However, the filter is only tested in experiments with a novel, univariate model.

An intriguing approach is that of \citet{stroud2007sequential}.
They peg the scaling of $\R$ and $\Q$ together,
so as to estimate only a single parameter, $\beta$.
This is difficult to justify, but simplifies the problem significantly
because then $\barB$ also scales with $\beta$ (provided linear dynamics)
so that $\pdf(\x,\beta) = \NormDist(\x|\bx,\beta\barB) \InvCS(\beta|\ldots)$
is conjugate to $\pdf(\y|\x,\beta)$,
and the full, joint posterior is available in closed form,
with trivial parametric updates,
hence avoiding Monte-Carlo.
However, without pegging $\Q$ to $\R$, the conjugacy is lost.

The EnKF-$N$ can also be said to be hierarchical
because of its careful treatment of the full marginal, $\pdf(\alpha|\y)$,
before the variational approximation \labelcref{eqn:post_EB}.

\section{More on the marginal inflation estimators}
\label{sec:more_on_marg}

This section complements \cref{sec:Marginal_s_estim}.

\subsection{Other trace-based estimators}
\label{sec:Other_variants}



It is common to form chi-square diagnostics
by measuring $\bbdelta$ by its Mahalanobis norm
\citep{menard2000assimilation,wu2013hyperparameter,haussaire2017thesis}.
This provides the motivation to use $\barC(1)^{-1}$
to transform \cref{eqn:expected_innov_3}, yielding:
\begin{align}
	\hat{\beta}_\barC =
	\frac{\norm{\bbdelta}^2_{\barC(1)} - \trace\big(\R\barC(1)^{-1}\big)}
	{\trace\big(\bH \barB \bH\tr \barC(1)^{-1}\big)}
	\, .
	\label{eqn:s_estim_C}
\end{align}

Alternatively, \cref{eqn:expected_innov_3} can be transformed
by $(\bH \barB \bH\tr)^{-1}$,
yielding
$\big( \bbdelta \bbdelta\tr - \R \big) (\bH \barB \bH\tr)^{-1}
=
\beta \I_P$
For the purpose of inflation,
this seems like the best option because then the trace
consists of terms with the same expected magnitude,
yielding the lowest aggregate variance.
The estimator becomes:
\begin{align}
	\hat{\beta}_{\bH\barB\bH\tr}
	&=
	\frac{1}{P}
	\big\{
		\norm{\bbdelta}^2_{\bH \barB \bH\tr}
		-\trace\big(\R (\bH \barB \bH\tr)^{-1}\big)
		\big\}
	\, ,
	\label{eqn:s_estim_HBH}
\end{align}
If $\bH \barB \bH\tr$ is rank-deficient,
then $\hat{\beta}_{\bH\barB\bH\tr}$ must be defined using the pseudo-inverse.
The estimate would then not be impacted by components of $\bbdelta$ outside of the ensemble subspace.

Lastly, note that none of the trace-based estimators are computationally costly,
because they can all be computed via the SVD \labelcref{eqn:svd_nRY}.

\subsection{Single-cycle bias}
\label{sec:Single_cycle_bias}

This subsection considers
the properties of the estimators within a single analysis,
based only on the inflation estimators' sampling distributions
Here, the assumption of \cref{sec:lit_review_0} that $N=\infty$ is undone,
so that $\barB$ is also random (before conditioning on the ensemble), in addition to $\x$, and $\y$,
and subject to sampling errors.
As will be shown, this
yields biases in the inflation estimators.

Now, using variables defined via the SVD \labelcref{eqn:svd_nRY},
it can be shown that each of the following estimators of $\beta$
satisfy the condition:
\begin{align}
	0 =
	\sum_{i=1}^P \gamma_i
	\big( 1 + \hat{\beta} \bar{\sigma}_i^2 - d_i^2 \big)
	\, ,
	\label{eqn:s_conds_0}
\end{align}
where
$d_i$ is the $i$-th component of the transformed innovation,
$\U\tr \R\msq \bbdelta$,
and
\begin{align}
	\gamma_i =
	\begin{cases}
		1                                & \text{for $\hat{\beta}_\R $}\\
		(1+\bar{\sigma}_i^2)^{-1}        & \text{for $\hat{\beta}_\barC $} \\
		(\bar{\sigma}_i^2)^{-1} 				 & \text{for $\hat{\beta}_{\bH\barB\bH\tr} $} \\
		(1+\bar{\sigma}_i^2)/(1+\hat{\beta} \bar{\sigma}_i^2)^2 & \text{for $\hat{\beta}_\tn{ML}$} \, . \\
	\end{cases}
	\label{eqn:s_conds}
\end{align}
\Cref{eqn:s_conds_0} may be solved explicitly for $\hat{\beta}$,
except in the case of $\hat{\beta}_\tn{ML}$.
Nevertheless, \cref{eqn:s_conds_0,eqn:s_conds} may be used to provide an insight on the bias of $\hat{\beta}_\tn{ML}$,
a subject of study since \citet{mitchell2000adaptive}.

Indeed, note that while the full matrix, $\barB$, is an unbiased estimator of $\B$,
the spectrum of $\barB$ is a biased estimate of the spectrum of $\B$
\citep{van1961certain,takemura1984orthogonally}.
Thus, the spectrum of $\bH \barB \bH\tr \R^{-1}$,
namely $\{\bar{\sigma}_i^2\}_{i=1}^P$, is also biased.
Hence, generally, functions of the spectrum will be biased.
An important exception is that
$\Expect(\sum_i \bar{\sigma}_i^2) = \trace(\bH \B \bH\tr \R^{-1})$,
meaning that the expectation of \cref{eqn:s_conds_0} holds for $\hat{\beta}_{\R}$.
Note that $\hat{\beta}_{\R}$ is still biased, as it requires inverting $\sum_i \bar{\sigma}_i^2$.
Nevertheless, considering the expressions \labelcref{eqn:s_conds} for $\gamma_i$,
it seems logical that the bias of $\{\bar{\sigma}_i^2\}_{i=1}^P$
will significantly carry over into the more complicated ones, such as $\hat{\beta}_\tn{ML}$.
Numerical experiments confirm this,
and show that the bias of $\hat{\beta}_\tn{ML}$ is worse than it is for $\hat{\beta}_\R $,
but less than for $\hat{\beta}_\barC $ and $\hat{\beta}_{\bH\barB\bH\tr} $.

Why is the bias of $\hat{\beta}_{\R}$ of \cref{eqn:s_miyoshi11} the least?
Loosely speaking, because the trace of $\barB$ is taken before dividing.
Moreover, in case $\R$ and $\bH \B \bH\tr$ have the same structure,
i.e. $\R^{-1/2} \bH \B \bH\tr \R\mtrsqrt = \sigma^2 \I_P$,
the bias of $\hat{\beta}_{\R}$ can be obtained analytically, as follows.
Due to the modified assumption \labelcref{eqn:fawefhawe},
\cref{eqn:awbuhawef} now yields $\barB \sim \Wishart(\B/\beta,\compactN)$.
Thus, the diagonal elements of $\R^{-1/2} \bH \barB \bH\tr \R\mtrsqrt$ are \iid{},
with distribution $\CS(\sigma^2/\beta,N{-}1)$.
Taking the trace increases the certainty to $\nu = P \cN$:
\begin{align}
	\bar{\sigma}^2 = \trace(\R^{-1/2}\bH \barB \bH\tr \R\mtrsqrt) \sim \CS(\sigma^2/\beta,\nu)
	\, .
\end{align}
Then, according to \cref{prop:CS} and \cref{tab:pdfs},
$\Expect [1/\bar{\sigma}^2] = \frac{\nu}{\nu-2}\beta/\sigma^2$.
Meanwhile,
$\Expect \big[ \bbdelta \bbdelta\tr \big] = \R + (1 + 1/\beta N)\bH \B \bH\tr$
so that
$\Expect \big[ \norm{\bbdelta}^2_\R/P - 1 \big] = (1 + 1/\beta N) \sigma^2$.
Now, the sample mean and variance of Gaussian samples are independent.
Hence, the nominator and denominator of $\hat{\beta}_{\R}$
of \cref{eqn:s_miyoshi11} are independent, so that
\begin{align}
	\Expect\big[\hat{\beta}_{\R}\big]
	&= \Expect \big[ \norm{\bbdelta}^2_\R/P - 1 \big] \Expect [1/\bar{\sigma}^2]
	\notag
	\\
	&= (1 + 1/\beta N) \frac{\nu}{\nu-2}\beta
	\, ,
\end{align}
which is close to $\beta$ for large $N$ and $P$.

\subsection*{Acknowledgements}
The authors thank
the two anonymous reviewers for their constructive comments;
A. Farchi for technical help and discussions;
L. Bertino for fostering the collaboration and for his patience;
C. Grudzien for insight on the behaviour of the Lorenz system;
F. Counillon, for reading and opinion;
A. Karspeck and M. Gharamti for stimulating questions.

Author P. N. Raanes has been funded by the EmblAUS project of the Nordic countries funding agency NordForsk,
and by DIGIRES, a project sponsored by PETROMAKS2 of the Research Council of Norway and industry partners.
CEREA is thanked for providing office space and community for visiting researcher P. N. Raanes. 
CEREA is a member of the Institut Pierre-Simon Laplace (IPSL).
Author A. Carrassi has been partly funded by the project REDDA of the Norwegian Research Council. 

\bibliographystyle{wileyqj}
\bibliography{../Refs/references}

\end{document}